\documentclass[aps,pre,twocolumn,groupedaddress,eqsecnum,showpacs,amsmath,amssymb]{revtex4}

\usepackage{graphicx}
\usepackage{epstopdf}

\newcommand{\ud}{\mathrm{d}}
\newcommand{\thetab}{\bar{\theta}}

\begin{document}

\title {Conformation of single-stranded RNA in a virus capsid: implications of dimensional reduction}

\author{Rouzbeh Ghafouri}
\email{rouzbeh@physics.ucla.edu}

\author{Joseph Rudnick}
\email{jrudnick@physics.ucla.edu}

\author{Robijn Bruinsma}

\email{bruinsma@physics.ucla.edu}

\affiliation{Department of Physics and Astronomy, UCLA, Los Angeles, CA 90095-1547}

\date{\today}

\begin{abstract}
The statistical mechanics of a treelike polymer in a confining volume is relevant to the packaging of the genome in RNA viruses. Making use of the mapping of the grand partition function of this system onto the statistical mechanics of a hard-core gas in two fewer spatial dimensions and of techniques developed for the evaluation of the equilibrium properties of a one-dimensional hard rod gas, we show how it is possible to determine the density and other key properties of a collection of rooted excluded-volume tress confined between two walls, both in the absence and in the presence of a one-dimensional external potential. We find, somewhat surprisingly, that in the case of key quantities, the statistical mechanics of the excluded volume, randomly branched polymer map exactly into corresponding problems for an \emph{unrestricted} linear polymer.
\end{abstract}


\maketitle

\section{Introduction and Motivation} \label{sec:intro}

Unlike the structure of the protein envelope of viruses, which is well-studied and precisely characterized \cite{crick,flint,baker} important aspects of the precise physical organization of the packaged genome are as yet undetermined. This is true in the case of both DNA \cite{cerritelli} and RNA  \cite{rao} viruses.  However, it has also been determined that a single strand of RNA will organize into a tree-like secondary structure \cite{fox,zuker}. Such secondary RNA structure occurs in general, and tree-like configurations are known to characterize the genomic conformation of certain RNA viruses \cite{brown, lu}. In light of those facts, one can hope to construct a reasonably accurate theoretical model of the statistical and mechanical properties of long segments of single-stranded (ss)  RNA---and in particular of the genomic matter in ss RNA viruses---if one can properly evaluate the statistical mechanics of a tree-like polymer in the presence of an external potential energy. The potential energy plays two roles in the context of the genomic conformation in RNA viruses. First, one naturally posits an energetic barrier that serves to confine the polymer to a particular region in space. Second, given the known interaction between RNA and the protein shell \cite{nagai,Morrell06011987}, it is reasonable to assume an attractive potential in the vicinity of the boundaries of that region.

The key challenge in this problem is taking into account the effects of excluded volume. That excluded volume effects are of central importance has been known since the late 1940's when Zimm and Stockmayer \cite{zs}  showed that the radius of gyration $R(N)$ of an ideal branching polymer scales with the number of monomers $N$ as $N^{1/4}$. This scaling relation means that non-excluded-volume branching polymers are highly condensed; the monomer density grows linearly with distance from the center of the polymer coil, a situation that cannot be sustained in light of excluded volume constraints. A field theory for excluded-volume interactions of branched polymers has been constructed by Lubensky and Isaacson  in the form of a $6-\epsilon$ expansion \cite{li}. More recently,  Parisi and Sourlas  \cite{ps} have utilized supersymmetry techniques to argue that the exponents of a $d$-dimensional branched polymer with excluded-volume interaction described by this field theory can be obtained by a mapping to the Yang-Lee edge singularity \cite{fish1,fish2} of a $d-2$ dimensional Ising model. For $d = 3$, this leads to the scaling relation $R(N) \propto N^{1/2}$ , with a density profile that now decreases inversely proportional to distance. The supersymmetry method is a demanding formalism, but Brydges and Imbrie \cite{bi1,bi2} showed that it could be reformulated as a relation between the conformational statistics of a branched polymer with excluded volume effects in d dimensions and the statistical mechanics of a hard-core liquid in $ d - 2$ dimensions. See also Cardy \cite{cardy1} for a particularly accessible exposition.

A previous paper by the present authors \cite{grb} contains an account of the utilization of dimensional reduction methods introduced by Brydges and Imbrie to determine the conformational properties, in particular the density, of rooted trees confined to a finite region and subject to an external potential. Although dimensional reduction holds in a curved geometry (see appendix \ref{app:curved}), such as the interior of a sphere---the geometry most relevant to RNA encapsulation in a viral capsid---it does not appear to lead to the kind of fundamental simplification that allows for the analysis of the effects of interactions between the branched polymer and the surrounding walls.  Consequently, our attention was focused on the simpler, but still relevant, problem of a branched polymer confined between two walls parallel in $3$ dimensions. We were able to allow for interactions between the polymer and the walls, as well as any other one-dimensional external potential energy. This paper provides background to that shorter work by filling in important calculational details. It also extends the results reported there. In particular, we build on the central--and somewhat surprising---outcome that the statistical mechanics of the excluded-volume branched polymer maps onto the statistics of an unrestricted chain polymer to  investigate both the density profile of the branched polymer and the interaction between bounding surfaces mediated by it. Given the substantial history of research on the statistics of unrestricted chain polymers in the literature(see, for instance, \cite{cai}), the results we present here are not entirely new.  However, we hope that they will prove stimulating in further investigations of the structure and assembly process of spherical viruses.

An outline of the paper is as follows. In Section \ref{sec:dimred} we review the consequences of the connection between the statistics of an excluded volume randomly branched polymer and the statistical mechanics of a gas of hard rods in two fewer dimensions. We focus on the simplest case of a one-dimensional hard rod gas, namely a gas in which the external potential is equal to zero. This is in order to develop key formulas and, additionally, to build some mathematical intuition with regard to the behavior of the more general system in which the external potential is not constant. Section \ref{sec:onedgas} addresses the means of solving for the density and partition function of the one-dimensional gas when the external potential varies spatially. The method utilized is based on an integral equation for the density of the gas derived to Percus \cite{percus}. We exploit the reformulation of that method by Vanderlick, \emph{et. al.} \cite{vanderlick}, which we recast into a form suitable for a ``lattice gas'' of one-dimensional hard rods. In Section \ref{sec:newmethod} the approach is further developed, so that the density of the gas, and hence the generating function for branched polymer statistics, follows from the solution of linear recursion relations. The connection between those recursion relations and the Schr\"{o}dinger equation is developed in Section \ref{sec:sch}. This connection provides the justification for the close relationship between the statistics of the excluded volume, randomly branched polymer and the unrestricted linear polymer chain.

\section{Dimensional reduction: the map from the tree-like polymer in $d$ dimensions to the hard-rod gas in $d-2$ dimensions} \label{sec:dimred}

The principal result that we will utilize relates the density, $n(x,z)$ of a hard-core gas in $d-2$ dimensions to the generating function, $\Sigma(x,z)$ of rooted, branched polymers in $d$ dimensions. Here, we assume translational symmetry in all but one dimension of the two systems, and $x$ is the one coordinate in the direction along which there is any spatial variation. The link between the two quantities is expressed in the two relationships \cite{bi1,bi2,cardy1}
\begin{eqnarray}
\rho_{d-2}(x,z) & = & \sum_{N=1}^{\infty}z(-z/\pi)^{N-1} N \mathcal{Z}_N (x)\label{eq:bi1} \\
\Sigma_{d}(x,z) & = &  \sum_{N=1}^{\infty} z^N\mathcal{Z}_N(x) \label{eq:bi2}
\end{eqnarray}
which tells us that the quantity $\mathcal{Z}_N$ contains information concerning the number of configurations of an $N$-monomer branched polymer  in $d$ dimensions that is rooted at the position $x$ (through (\ref{eq:bi2})) or, alternatively, concerning the density at $x$ of an $N$-particle gas with hard core repulsion in $d-2$ dimensions (through (\ref{eq:bi1})). According to the above equations, the behavior of the polymer system with positive fugacity, $z$, is obtained by investigating the mathematical structure of the gas in the grand canonical ensemble, but with \emph{negative} fugacity.

As an example of the application of Eqs (\ref{eq:bi1}) and (\ref{eq:bi2}), and to establish some points of reference for the discussion to follow, we will review the statistical mechanics of a gas of rods in a very large one-dimensional interval subject to a constant potential energy.

\subsection{One-dimensional rods in an extended region under the influence of a constant external potential} \label{sec:prelim}

The constant external potential can be set equal to zero. The grand partition function of the one-dimensional gas is given by
\begin{equation}
\mathcal{Q}(L) = \sum_{N=0}^{\infty} \frac{(L-aN)^N}{a^N N!} z^{N} \ \ \ \ N<La
\label{eq:examp1}
\end{equation}
where $a$ is the hard-core ``radius.'' The factor $a^N$ in the denominator guarantees that each term in the summand is dimensionless. In the limit of very large $L$, the sum will be dominated by the $N$ value for which the summand is maximum. To locate that term we first exponentiate the summand and then re-express the exponent in terms of the density $\rho=N/L$. Then the summand is of the form
\begin{equation}
\exp \left[L \left( \rho \ln \frac{1-a\rho}{a \rho} + \rho + \rho \ln z \right) \right]
\label{eq:exsummand1}
\end{equation}
where we have made use of Stirling's formula: $\ln N \approx N\ln(N/e)$. The extremum equation that follows from an attempt to maximize (\ref{eq:exsummand1}) with respect to $N$, and hence $\rho$, is
\begin{equation}
\ln \frac{1-a \rho}{a \rho} - \frac{a \rho}{1- a \rho} + \ln z =0
\label{eq:exext1}
\end{equation}

\begin{figure}[htbp]
\begin{center}
\includegraphics[width=3in]{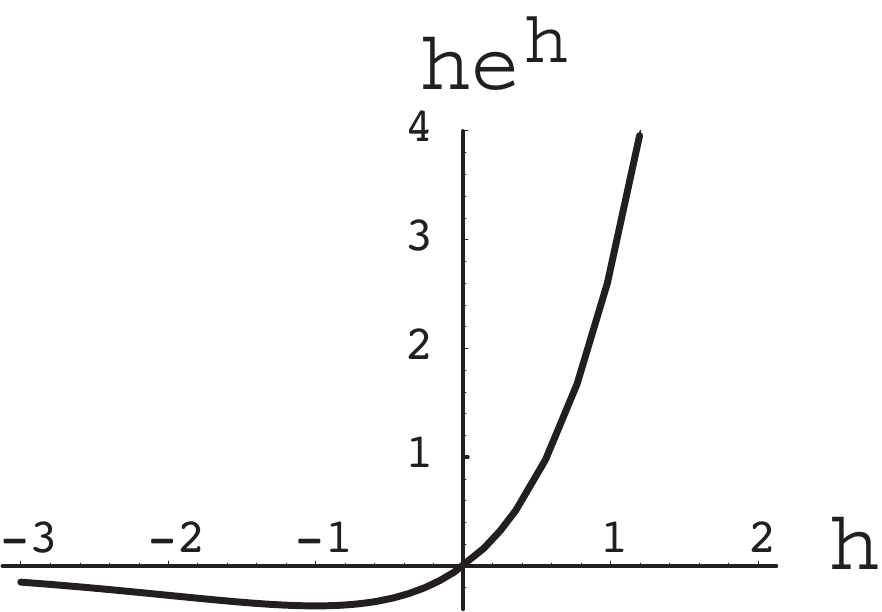}
\caption{The function, $he^h$, on the left hand side of (\ref{eq:exext3}), graphed as a function of $h$.}
\label{fig:lhs1}
\end{center}
\end{figure}

\begin{figure}[htbp]
\begin{center}
\includegraphics[width=3in]{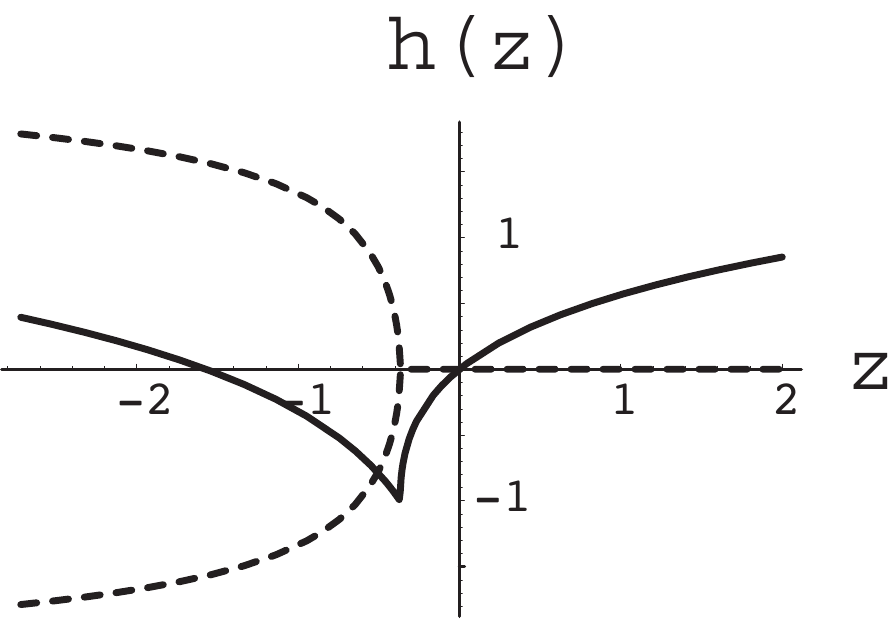}
\caption{The real (solid curve) and imaginary (dashed curves) parts of solutions to (\ref{eq:exext3}).}
\label{fig:Wfun1}
\end{center}
\end{figure}

We introduce the new variable
\begin{equation}
h = \frac{a \rho}{1-a \rho}
\label{eq:exh1}
\end{equation}
Then,
\begin{equation}
\rho = \frac{1}{a}\frac{h}{1+h}
\label{eq:exh2}
\end{equation}
and (\ref{eq:exext1}) becomes
\begin{equation}
- \ln h - h+ \ln z =0
\label{eq:exext2}
\end{equation}
or
\begin{equation}
he^h = z
\label{eq:exext3}
\end{equation}
The function $he^h$ on the left hand side of (\ref{eq:exext3}) is graphed in Fig. \ref{fig:lhs1}.
The solution to Eq. (\ref{eq:exext3}) is, formally, $h=W(z)$, where $W$ is the Lambert $W$ function \cite{corless}. Figure \ref{fig:Wfun1} shows what the solution looks like, as the fugacity varies from a negative to positive values. Note the onset of imaginary components to the solution. The departure from a purely real solution follows from the minimum in the function $he^h$, as displayed in Fig. \ref{fig:lhs1}. When $z$ lies below this minimum, there is no purely real solution to (\ref{eq:exext3}).

The behavior of the solutions in the vicinity of the point at which the imaginary part emerges yields information about both the location of this ``critical point'' and about its implications for the statistics of self-avoiding rooted trees in three dimensions. The minimum in $he^{h}$ is at $h=h_c=-1$, corresponding to $z=z_c =-1/e$ in (\ref{eq:exext3}). Expanding both sides of that equation about those critical values we have
\begin{eqnarray}
(h_c + \delta h)e^{h_c+\delta h} & = & -\frac{1}{e} + \frac{\delta h^2}{2e} + O(\delta h^3) \nonumber \\
& = & z_c + \delta z \nonumber \\
& = & -\frac{1}{e} + \delta z
\label{eq:criteq1}
\end{eqnarray}
Solving to lowest order for $\delta h$:
\begin{equation}
\delta h =  \pm \sqrt{2e \delta z}
\label{eq:criteq2}
\end{equation}
This tells us that, in the immediate vicinity of $z=z_c$,
\begin{equation}
h=-1\pm \sqrt{2e \delta z}
\label{eq:criteq3}
\end{equation}
or, from (\ref{eq:exh2})
\begin{equation}
\rho \rightarrow \mp \frac{1}{a}\frac{1}{\sqrt{2e \delta z}}
\label{eq:criteq4}
\end{equation}
When $ z > z_c$ continuity of the solution with $z>0$ requires that we take the upper sign in (\ref{eq:criteq2})--(\ref{eq:criteq4}). However, when $z<z_c$, the choice of sign is controlled by the choice of location with respect to a branch cut in the complex $z$ plane, starting at the branch cut at $z=z_c$ and extending to $z=-\infty$ along the negative $z$ axis.

\begin{figure}[htbp]
\begin{center}
\includegraphics[width=3in]{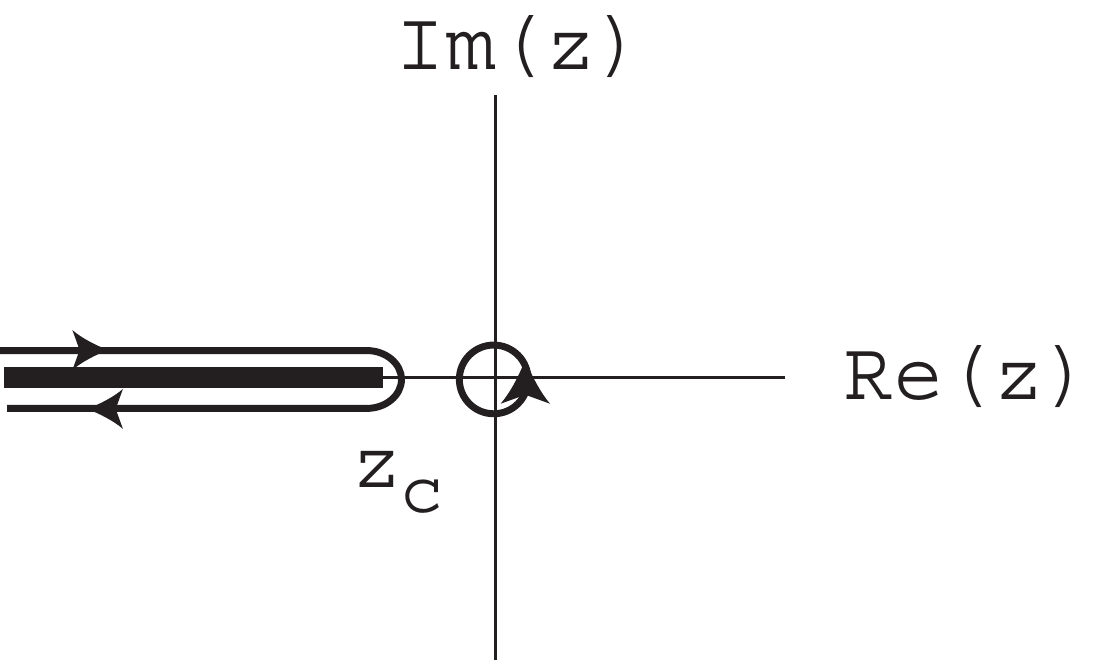}
\caption{The original contour in (\ref{eq:criteq5}) (small circle surrounding the origin) and the distorted contour that bounds the branch cut originating at the branch point, $z_c$. }
\label{fig:distort}
\end{center}
\end{figure}

The fact that the (uniform) density of the one-dimensional hard rod gas possess a singularity going as $(z-z_c)^{-1/2}$, where $z_c=-1/e$, allows us to extract the leading behavior of the large $N$ coefficients in a power series expansion of this function of the fugacity \cite{rudgas}. Making use of the general result that if a function $f(z)$ can be written as  a power series about $z=0$, so that $f(z) = \sum_{N=0}^{\infty} F_N z^N$, then
\begin{equation}
F_N = \frac{1}{2 \pi i} \oint \frac{f(z)}{z^{N+1}} dz
\label{eq:criteq5}
\end{equation}
where the contour over which the integral in (\ref{eq:criteq5}) is performed encircles the origin in the complex $z$ plane and does not enclose any singularities in the function $f(z)$. Expanding the contour so that it impinges on the branch point and, ultimately surrounds the branch cut (see Fig. \ref{fig:distort}), we end up with the following result for the coefficient of $z^N$ in the power series expansion of $\rho(z)$:
\begin{eqnarray}
\lefteqn{\frac{1}{\pi a \sqrt{2e}} \int_0^{\infty}\frac{(-1)^{N+1}}{(+|z_c|+ \delta z)^{N+1}} \delta z^{-1/2} \  d \delta z} \nonumber \\ & = & \frac{1}{\pi a \sqrt{2e}} \frac{(-1)^{N+1}}{|z_c|^{N+1}} \int _0^{\infty} e^{-(N+1) \ln (1+ \delta z/|z_c|)} \delta z^{-1/2}  \ d \delta z \nonumber \\
& \rightarrow& \frac{1}{\pi a \sqrt{2e}} \frac{(-1)^{N+1}}{|z_c|^{N+1}}\int_0^{\infty} e^{-(N+1) \delta z/|z_c|} \delta z^{-1/2}  \ d \delta z \nonumber \\
& \propto & \frac{(-1)^{N+1}}{|z_c|^{N+1/2}}N^{-1/2} \ \ \ \ (N \gg 1)
\label{eq:criteq6}
\end{eqnarray}
According to (\ref{eq:bi1}) and (\ref{eq:bi2}), we see that the number of conformations of three dimensional trees with $N$ monomers should grow geometrically as $1/(\pi z_c)^{N}$, with the additional power law modification $N^{-3/2}$.

Note that in the above discussion, we have neglected the possibility of any other non-analyticity in $\rho(z)$. As one can readily verify, the leading contributions to the integral in (\ref{eq:criteq5}) will, for large $N$, be controlled by the singularities in $\rho(z)$ that lie closest to the origin. The possible existence of other poles, branch points or essential singularities in $\rho(z)$, all of which will occur at $|z| > |z_c|$, is irrelevant to the results obtained here.

The fact that the statistics of the hard-core gas---and by extension randomly configured trees---are controlled by non-analyticity in the grand canonical ensemble that is closest to the origin in the complex $z$ plane can be exploited to infer the emergence of a bound state generated by the presence of an attractive external potential. The argument is as follows \cite{cardy1}.

We start with the density at a particular point of the gas of hard rods. In the grand canonical ensemble, this density equal to the sum over configurations in which the a rod is at that point, divided by the sum over all allowable configurations. If the system is confined to the one dimensional region between $x=0$ and $x=L$, then this leads to
\begin{equation}
\rho(x,z) = \frac{Q(x,z) z Q(L-x-a)}{Q(L,z)}
\label{eq:denspot1}
\end{equation}
with $Q$ the grand partition function as before. If, now, there is a non-zero, delta function potential energy, $V$ at the position $x_1$, then the grand partition function contains an additional contribution from the associated Boltzmann factor weighting configurations in which a gas particle occupies that particular location. If we write
\begin{equation}
e^{- \beta V} = 1 + \lambda
\label{eq:lamdef}
\end{equation}
where an attractive potential (negative $V$) generates a $\lambda >0$, then the new partition function is
\begin{eqnarray}
Q^{\prime}(L,z)  &=& Q(L,z) + Q(x_1,z) \lambda z Q(L-x_1,z) \nonumber \\
& = & Q(L,z) (1+ \lambda  \rho(x_1,z))
\label{eq:newQ}
\end{eqnarray}

The new density at $x_1$ is, then, given by
\begin{eqnarray}
\rho^{\prime}(x_1,z) &=& \frac{Q(x_1,z) z Q(L-x_1,z)}{Q^{\prime}(L,z)} \nonumber \\
& = & \frac{Q(x_1,z)  z Q(L-x_1,z)}{Q(L,z)(1+\lambda \rho(x_1,z))}
\label{eq:newnp}
\end{eqnarray}

\begin{figure}[htbp]
\begin{center}
\includegraphics[width=3in]{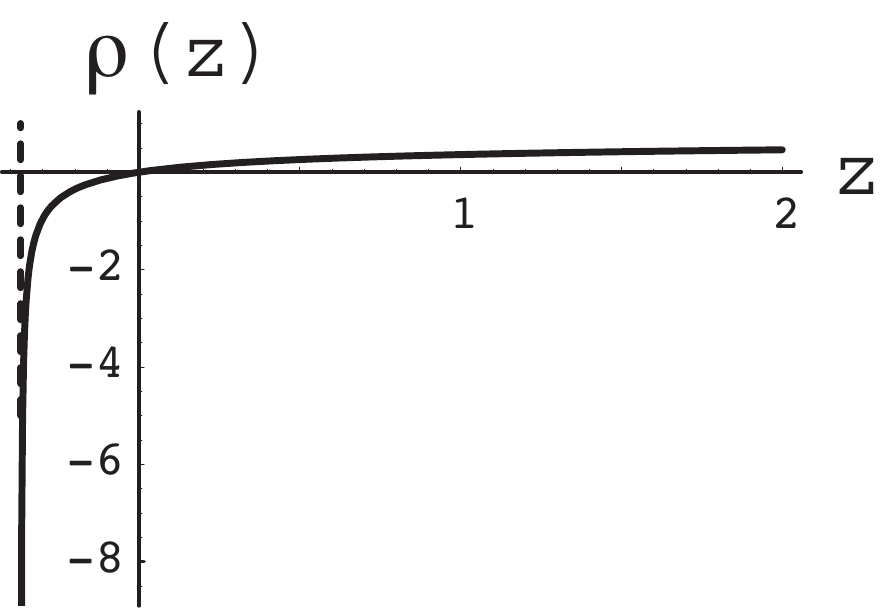}
\caption{The function $\rho(z)$, as given by (\ref{eq:exext1})--(\ref{eq:exext3}), for $z>z_c$, the location of which is indicated by the vertical dashed line. }
\label{fig:nplot}
\end{center}
\end{figure}

In the case of a very long interval and a large value of $x_1$, the density $\rho(x_1, z)$ will be independent of $x_1$, and is  given by (\ref{eq:exext1})--(\ref{eq:exext3}). Figure \ref{fig:nplot} illustrates the behavior of the density as a function of $z$ for $z>z_c=-1/e$.
The fact that $\rho(z)$ is negative for $z<0$ and that it goes to $-\infty$ as $z\rightarrow z_c^{+}$ (see also (\ref{eq:criteq4})) ensures that, for any positive value of $\lambda$, there will be a zero in $Q^{\prime}(L.z)$ and hence a pole in $ \rho^{\prime}(x_1,z)$ for a (negative) value of $z$ closer to the origin than $z_c$. If we associate this singularity in the density with a bound state, we are led to conclude that an arbitrarily weak one-dimensional potential well in the interior of an infinitely extended system will ``bind'' a three dimensional random tree. Further analysis of this situation requires a more searching exploration of the spatial structure of this putative bound state.

\section{One dimensional hard core gas in the presence of an external potential. } \label{sec:onedgas}

The problem of a one dimensional gas of hard core rods was considered by Percus \cite{percus}, who derived an integral equation for the equilibrium density of that system, from which quantity the partition function and all interesting equilibrium correlation functions can be obtained. The integral equation, which has been reduced to a very useful and tractable form by Vanderlick, \emph{et. al.} \cite{vanderlick}, serves as a starting point for the exploration of the equilibrium statistics of randomly branched polymers in a one-dimensional environment, given the connections established by Brydges and Imbrie \cite{bi1,bi2}. The two equations leading to the calculation of the density in this one-dimensional system are \cite{percus,vanderlick}
\begin{eqnarray}
	h(x)& = & \frac{\rho(x)}{1-\int_{x}^{x+\sigma}{\rho(t)dt}} \label{eq:van1:nh} \\
	h(x)&=&e^{\beta(\mu-u(x))} \exp \left[-\int_{x-\sigma}^{x}{h(t)dt} \right] \label{eq:van1:hh}
\end{eqnarray}
where (\ref{eq:van1:nh}) can be taken as a definition of the function $h(x)$.

Note the strong similarity between (\ref{eq:van1:hh}) and (\ref{eq:exext3}). In fact, if we set $u(x)=0$, replace $e^{\beta \mu}$ by $z$, take $\sigma=1$, and assume an $x$-independent $h(x)$, then (\ref{eq:van1:hh}) reproduces (\ref{eq:exext3}), while, if we also assume an $x$-independent density, $\rho$, and set $a=1$, then (\ref{eq:van1:nh}) reduces to (\ref{eq:exh1}).

For our purposes, it proves more useful to focus our investigations on a discrete version of the one-dimensional gas of rods---a one-dimensional, hard core lattice gas. As it turns out, the equations leading to a solution for the density of this system have already been worked out in a different context \cite{schwab}. To maintain a self-contained exposition, the derivation of the equations governing the density in this discrete system are also presented below. This  derivation closely parallels the arguments of Percus \cite{percus}.

\begin{figure}[htbp]
\begin{center}
\includegraphics[width=3in]{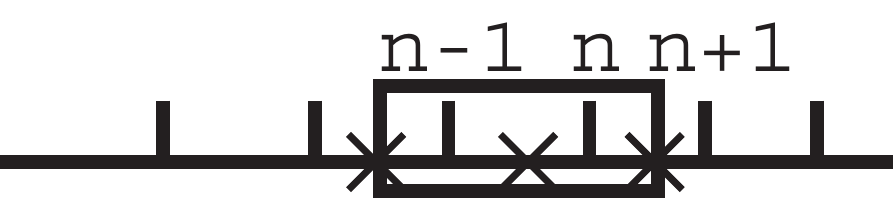}
\caption{A rod with a lenth $\sigma=2$ located at the site $n$. }
\label{fig:rodloc}
\end{center}
\end{figure}

We assume rods that can sit on specific points on a line.
\begin{figure}[htbp]
\begin{center}
\includegraphics[width=3in]{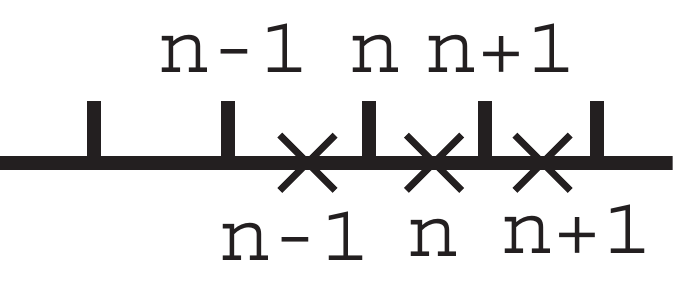}
\caption{The setup for the one dimensional hard rod gas.}
\label{fig:onedgas}
\end{center}
\end{figure}
As shown in Fig. \ref{fig:onedgas}, the possible locations for the rods are at sites labeled by the integer $n$. The actual locations of the sites are indicated by the vertical lines in the figure. The boundaries between different regions of the gas lie between the sites at the points indicated by the $\times$'s. These boundaries are utilized in the definitions of ``partial'' grand partition functions. For instance the grand partition function $\Xi(-\infty, n)$ corresponds to a system in which the rods can occupy the region to the left of the $\times$ at $n$ in Fig. \ref{fig:onedgas}. The full grand partition function of the infinite system is $\Xi(- \infty, \infty) \equiv \Xi$. Now, assume a chemical potential $\mu$ and site-dependent local potentials $u_n$. We locate a rod by indicating the site at which its far right portion sits. For instance, Fig. \ref{fig:rodloc} shows a rod with a length equal to two that is located at the site $n$.
Given this, it is straightforward to show that the density of rods at the site $n$ is given by
\begin{equation}
\rho_n= \frac{\Xi(-\infty,n-\sigma) \Xi(n, \infty) e^{\beta(\mu-u_n)}}{\Xi}
\label{eq:denseq1}
\end{equation}

The next step is to establish a relation for the product of two partial grand partition functions. In particular, we are interested in $\Xi(-\infty, n) \Xi(n, \infty)$. This product is almost the entire partition function. In fact, it only omits configurations in which a rod overlaps the boundary at $n$. In the case of the rod pictured in Fig. \ref{fig:rodloc}, there is precisely one such configuration: the one in which the right hand side of the rod lies on the site $n+1$. In general there are $\sigma-1$ such configurations. Let us look at the one corresponding to the location of the rod with length $\sigma$ lying on the site $n+1$. The contribution to $\Xi$ of that particular configuration is
\begin{equation}
\Xi(-\infty,n+1-\sigma) \Xi(n+1, \infty) e^{\beta(\mu-u_{n+1})} = \Xi \rho_{n+1}
\label{eq:missingpart1}
\end{equation}

Making the appropriate corrections for all missing contributions to $\Xi$, we see that
\begin{equation}
\Xi(-\infty, n) \Xi(n, \infty) = \Xi \times \left(1- \sum_{j=1}^{\sigma-1} \rho_{n+j}\right)
\label{eq:sigprod}
\end{equation}

Now, we construct a recursion relation for the partial grand partition function. By inspection, one sees that the difference between the partial function $\Xi(-\infty,n)$ and the function $\Xi(-\infty,n-1)$ lies in configurations in which there is a rod at $n$. In other words
\begin{eqnarray}
\Xi(-\infty, n) & = & \Xi(-\infty,n-1) + e^{\beta(\mu-u_n)} \Xi(- \infty,n-\sigma) \nonumber \\
& = & \Xi( -\infty,n-1) + \frac{\rho_n \Xi}{\Xi( n, \infty)} \nonumber \\
& = & \Xi(-\infty,n-1) + \frac{\rho_n \Xi(-\infty, n)}{1- \sum_{j=1}^{\sigma-1} \rho_{n+j}}
\label{eq:recur1}
\end{eqnarray}
The second line of (\ref{eq:recur1}) follows from (\ref{eq:denseq1}) and the last line from (\ref{eq:sigprod}).
Thus,
\begin{eqnarray}
\Xi(-\infty, n-1) &=& \Xi(-\infty,n)\left[1- \frac{\rho_n}{1- \sum_{j=1}^{\sigma-1} \rho_{n+j}}\right] \nonumber \\
& \equiv & \Xi(-\infty, n)(1-h_n)
\label{eq:recur2}
\end{eqnarray}
where the last line of (\ref{eq:recur2}) serves as a definition of the quantity $h_n$, given as
\begin{equation}
h_n = \frac{\rho_n}{1- \sum_{j=1}^{\sigma-1} \rho_{n+j}}
\label{eq:hdef}
\end{equation}

Returning to the expression on the right hand side of (\ref{eq:denseq1}), we note that, given (\ref{eq:recur2}),
\begin{widetext}
\begin{eqnarray}
\rho_n   & = &e^{\beta(\mu-u_n)}  \frac{\Xi(-\infty,n) (1-h_n)(1-h_{n-1}) \cdots (1-h_{n-\sigma+1}) \Xi(n,\infty)}{\Xi} \nonumber \\
& = & e^{\beta(\mu-u_n)} (1-\sum_{j=1}^{\sigma-1} \rho_{n+j}) \prod_{m=1}^{\sigma}(1-h_{n-m+1})
\label{eq:rhofin1}
\end{eqnarray}
\end{widetext}
We then divide both sides of (\ref{eq:rhofin1}) by the term $1- \sum_{j=1}^{\sigma-1} \rho_{n+j}$. Then, making use of the definition of $h_k$ implicit in the last line of (\ref{eq:recur2}), and singling out the first term in the product on the right hand side of (\ref{eq:rhofin1}) we have
\begin{equation}
h_n = e^{\beta(\mu-u_n)} (1-h_n) \prod_{m=2}^{\sigma}(1-h_{n-m+1})
\label{eq:rhofin2}
\end{equation}
Solving for $h_n$, we are left with
\begin{equation}
h_n = \frac{e^{\beta(\mu-u_n)} \prod_{m=2}^{\sigma}(1-h_{n-m+1})}{1+e^{\beta(\mu-u_n)}\prod_{m=2}^{\sigma}(1-h_{n-m+1})}
\label{eq:rhofin3}
\end{equation}
If we define
\begin{equation}
H_n = \prod_{m=2}^{\sigma}(1-h_{n-m+1})
\label{eq:Hdef}
\end{equation}
then the equation (\ref{eq:rhofin3}) becomes
\begin{equation}
h_n = \frac{H_ne^{\beta(\mu-u_n)}}{1+H_n e^{\beta(\mu-u_n)}}
\label{eq:newheqn}
\end{equation}
The quantity $H_n$ can be thought of in terms of a local modification of the fugacity arising from the excluded volume constraint. Note that the quantity $h_n$ is determined by $h_n$'s to the left of it. This means that the system of equations (\ref{eq:rhofin3}) can be solved by iteration to the right. Furthermore, given the explicit definition of $h_n$ in equation \eqref{eq:hdef}, we see that the density, $\rho_n$ is expressed in terms of $h_n$ and $\rho_m$'s with $m >n$, so having solved for $h_n$, we obtain the $\rho_n$'s by iterating to the left.

To highlight the precise points of reference between the discrete equations above and the integral equations developed by Percus \cite{percus} and refined by Vanderlick, \emph{et. al.} \cite{vanderlick}, we note that (\ref{eq:hdef}) corresponds to (\ref{eq:van1:nh}) while (\ref{eq:rhofin3}) is the analogue in the discrete system of (\ref{eq:van1:hh}).

Another version of the set of equations that is a bit simpler to iterate replaces the variables $h_n$ by
\begin{equation}
g_n=1-h_n
\label{eq:gdef1}
\end{equation}
Then,
\begin{equation}
H_n = \prod_{m=2}^{\sigma} g_{n-m+1}
\label{eq:newHdef}
\end{equation}
and the equation for $g_n$ becomes
\begin{equation}
g_n = \frac{1}{1+H_n e^{\beta(\mu-u_n)}}
\label{eq:geq}
\end{equation}
with
\begin{equation}
g_n= 1-\frac{\rho_n}{1- \sum_{j=1}^{\sigma-1} \rho_{n+j}}
\label{eq:gdef}
\end{equation}
The equations above constitute discrete version of the method of Vanderlick, {\em et. al.} \cite{vanderlick}.

\subsection{The case $\sigma=1$}

It is worthwhile to ask what happens when $\sigma=1$. In this case, the rods are ``point particles,'' and excluded volume does not play a role. The solution of the equations is greatly simplified. Given (\ref{eq:rhofin1}) and (\ref{eq:hdef}) with $\sigma=1$, one arrives at the following end-result.
\begin{equation}
\rho_n = \frac{e^{\beta(\mu-u_n)}}{1+e^{\beta(\mu-u_n)}}
\label{eq:pprho}
\end{equation}
The lattice gas in this case is, in fact, well-described by an appropriate version of the one-dimensional Ising model. This system maps onto trees in which the ``branches'' have zero extension, and in which, furthermore, the radius of the hard core at the vertices between branches is vanishingly small.

\subsection{The lattice gas with $\sigma=2$} \label{sec:sigmatwo}

The simplest non-trivial version of this lattice gas as it applies to the conformational statistics or random trees has $\sigma=2$. Then (\ref{eq:rhofin3}) and (\ref{eq:hdef}) become
\begin{eqnarray}
h_n & = & \frac{e^{\beta(\mu-u_n)}(1-h_{n-1})}{1+e^{\beta(\mu-u_n)}(1-h_{n-1})} \label{eq:heq2} \\
h_n & = & \frac{\rho_n}{1-\rho_{n+1}} \label{eq:rhoeq1}
\end{eqnarray}

For the time being, we will focus on a uniform gas confined to a finite region. This means that we are going to set $u_n=0$, supplemented by boundary conditions to be developed below. Replacing $e^{\beta \mu}$ by the fugacity, $z$, the equation for $h_n$ becomes
\begin{equation}
h_n = \frac{z(1-h_{n-1})}{1+z(1-h_{n-1})}
\label{eq:heq3}
\end{equation}
In terms of $g_n$, as defined in (\ref{eq:gdef1}), (\ref{eq:heq3}) becomes
\begin{equation}
g_n = \frac{1}{1+zg_{n-1}}
\label{eq:geq1}
\end{equation}
This equation leads to a solution for $g_n$ in terms of $g_1$ in terms of continued fractions:
\begin{equation}
g_n=\frac{1}{1+\frac{z}{1+\frac{z}{1+ \cdots}}}
\label{eq:gsol1}
\end{equation}
We can rewrite the solution to the equation for $g_n$ in terms of an iterated matrix equation  \cite{wall}. Given the fact that the continued fraction is repeated, the iteration of the matrix equation is relatively straightforward. Suppose we have the solution for $g_m$ in terms of $g_{m-l}$ as follows:
\begin{equation}
g_m = \frac{A_l+B_l g_{m-l}}{C_l+D_lg_{m-l}}
\label{eq:gsol2}
\end{equation}
Then, given
\begin{equation}
g_{m-l} = \frac{1}{1+zg_{m-l-1}}
\label{eq:gsol3}
\end{equation}
we have
\begin{eqnarray}
g_l & = & \frac{A_{l+1}+B_{l+1}g_{m-l-1}}{C_{l+1}+D_{l+1}g_{m-l-1}} \nonumber \\
& = & \frac{A_l+B_l \frac{1}{1+zg_{m-l-1}}}{C_l +D_kl\frac{1}{1+zg_{m-l-1}}}
\label{eq:gsol4}
\end{eqnarray}
Rationalizing the last line of (\ref{eq:gsol4}), we end up with the following recursion relations
\begin{eqnarray}
A_{l+1} & = & A_l+B_l \label{eq:Aeq} \\
B_{l+1} & = & zA_l \label{eq:Beq} \\
C_{l+1} & = & C_l+D_l \label{eq:Ceq} \\
D_{l+1} & = & zB_l \label{eq:Deq}
\end{eqnarray}
Equations (\ref{eq:Aeq}) and (\ref{eq:Beq}) can be written as follows
\begin{equation}
\left( \begin{array}{c} A_{l+1} \\  B_{l+1} \end{array} \right) = \left( \begin{array}{cc} 1 & 1 \\ z & 0 \end{array} \right) \left( \begin{array}{c} A_l  \\  B_l \end{array} \right)
\label{eq:mat1}
\end{equation}
with a similar equations for $C$ and $D$. Given that we have the initial conditions $A_0=0$, $B_0=1$, $C_0=1$ and $D_0 =0$, we have
\begin{equation}
\left( \begin{array}{c} A_{l} \\  B_{l} \end{array} \right) = \left( \begin{array}{cc} 1 & 1 \\ z & 0 \end{array} \right)^l \left( \begin{array}{c} 0  \\  1 \end{array} \right)
\label{eq:mat2}
\end{equation}
Our task now is to take the $l^{\rm th}$ power of the matrix in (\ref{eq:mat1}).

\subsection{A digression: the uniform case}

\begin{figure}[htbp]
\begin{center}
\includegraphics[width=3in]{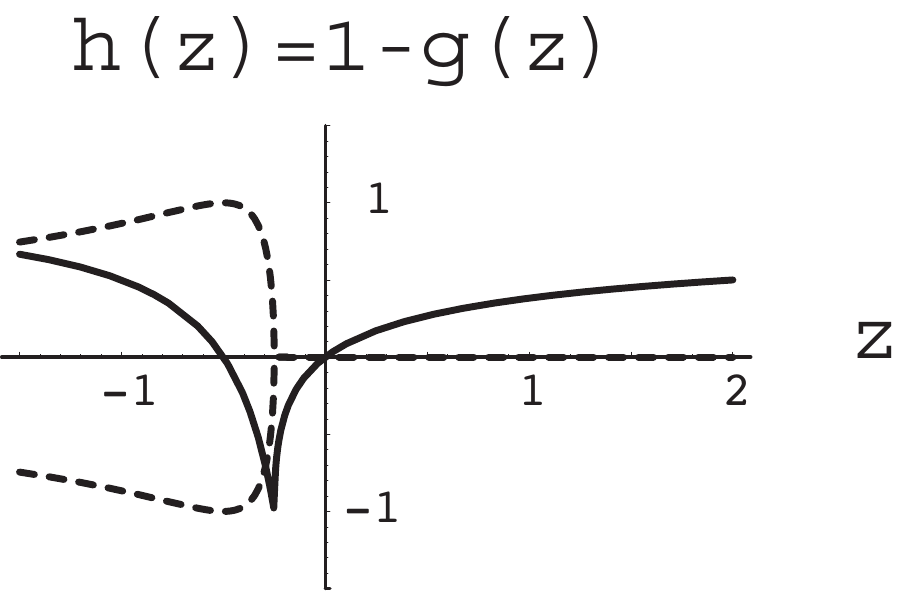}
\caption{The real (solid curve) and imaginary (dashed curves) parts of $h(z)=1-g(z)$, where $g(z)$ is as given by (\ref{eq:constg2}). }
\label{fig:gplot}
\end{center}
\end{figure}

Before performing the requisite calculations, we will consider the solution (\ref{eq:geq1}) under the assumption that the g's are independent of location. The equation that results from that model is
\begin{equation}
zg^2+g-1=0
\label{eq:constg1}
\end{equation}
The solution is
\begin{equation}
g=\frac{1}{2z} \left[\sqrt{1+4z}-1 \right]
\label{eq:constg2}
\end{equation}
There is a branch point at $z=-1/4$. Figure \ref{fig:gplot} shows what the real and imaginary parts of $g$ look like as a function of $z$.
Close to the critical point, the general form of the plot is qualitatively identical to the result for $h(z)$ displayed in Fig. \ref{fig:Wfun1}.

We can reconstruct the (constant) density, $\rho$, from (\ref{eq:rhoeq1}), with $h=1-g$, where $g$ is given by (\ref{eq:constg2}). From (\ref{eq:rhoeq1}) with $\rho$, $h$ and $g$ constant, we have
\begin{eqnarray}
\rho& = & \frac{h}{1+h} \nonumber \\
& = & \frac{1-g}{2-g} \nonumber \\
& = & \frac{2z+1-\sqrt{1+4z}}{1+4z - \sqrt{1+4z}}
\label{eq:constg3}
\end{eqnarray}
If $z=-1/4 + \delta$, then
\begin{eqnarray}
\rho & = & \frac{1/2 + 2 \delta-2\sqrt{\delta}}{4 \delta - 2 \sqrt{\delta}}  \nonumber \\
&\rightarrow &  - \frac{1}{4 \sqrt{\delta}}
\label{eq:constg4}
\end{eqnarray}

The solution (\ref{eq:constg2}) suggests a reparameterization that is useful in the vicinity of the ``critical point.''
\begin{equation}
z = - \frac{1}{4 \cosh^2k}
\label{eq:param1}
\end{equation}
Then, (\ref{eq:constg2}) becomes
\begin{equation}
g = 2 \cosh k e^{-k}
\label{eq:constg5}
\end{equation}
When $z$ passes through -1/4,  (\ref{eq:param1}) requires that we replace the real $k$ by an imaginary quantity.

\subsection{Completion of the calculation}
The results we need follow from the eigenvectors and eigenvalues of the matrix $\left( \begin{array}{cc} 1 & 1 \\ z & 0 \end{array} \right)$. Solving the relevant equations and
making use of the reparameterization (\ref{eq:param1}), we find that (\ref{eq:mat2}) reduces to
\begin{equation}
\left( \begin{array}{c} A_n \\ B_n \end{array} \right) = \left( \begin{array}{c} \frac{\sinh kn}{(2 \cosh k)^{n-1} \sinh k} \\ -\frac{\sinh(n-1)k}{(2 \cosh k)^n \sinh k} \end{array} \right)
\label{eq:matres2}
\end{equation}
With a similar set of steps, we end up with
\begin{equation}
\left( \begin{array}{c} C_n \\ D_n \end{array} \right) = \left( \begin{array}{c} \frac{\sinh (n+1) k}{\sinh k (2 \cosh k)^n} \\ - \frac{\sinh kn}{ \sinh k (2 \cosh k)^{n+1}} \end{array} \right)
\label{eq:matres3}
\end{equation}

\begin{figure}[htbp]
\begin{center}
\includegraphics[width=3in]{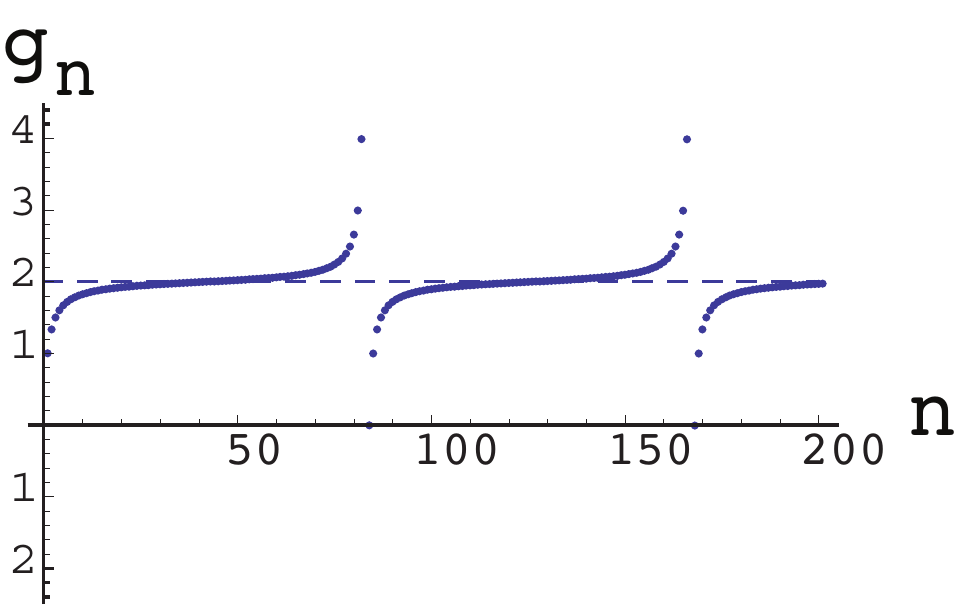}
\caption{The function $g_n$ when $z$ is slightly to the right of the critical point at -1/4. The dashed line at $g_n=2$ indicates the value to which $g_n$ tends when $z=-1/4$ exactly. }
\label{fig:gplot1}
\end{center}
\end{figure}

We assume that the initial $g$ is equal to zero. This is consistent with assumptions in the case of the ordinary one-dimensional excluded-volume gas. We will assign this $g$ the index 0. Then,
\begin{eqnarray}
g_n(z) &= &\frac{A_n}{C_n} \nonumber \\
& = & \frac{2 \sinh nk \cosh k}{\sinh (n+1) k} \nonumber \\
& = & 2 \cosh^2k - \sinh 2k \coth(n+1)k
\label{eq:ggen1}
\end{eqnarray}
If $k$ is real, the $n \rightarrow \infty$ limit of the above expression is consistent with (\ref{eq:constg3}). However, if $k$ is imaginary, then the second term on the last line of (\ref{eq:ggen1}) behaves like the cotangent function, and is periodic with a set of poles. See Fig. \ref{fig:gplot1}.
The graph in this figure was generated by numerical solution of the recursion relation (\ref{eq:geq1}). The key property of this plotted solution for $g_n$, and in the analytical form in (\ref{eq:ggen1}), is the appearance of poles. In fact, as will be demonstrated below, the passage of a pole in $g_n$ through the far boundary of the interval of interest is directly associated with the appearance of a singularity in the dependence of the density on the fugacity, $z$. A fuller discussion of this point will follow the development of an alternative approach to the solution of the recursion relation for $g_n(z)$.

\section{Modified approach to the solution of the recursion relations} \label{sec:newmethod}

We now describe an alternate route to the analysis of this system. Recall Eq. (\ref{eq:geq1}). Suppose we write
\begin{equation}
1+zg_n(z) = \frac{\psi_{n+1}(z)}{\psi_n(z)}
\label{eq:psidef}
\end{equation}
Then, (\ref{eq:geq1}) is transformed to the following recursion relation
\begin{equation}
\frac{\psi_{n+1}(z)}{\psi_n(z)} = 1 + \frac{z}{\psi_n(z)/\psi_{n-1}(z)}
\label{eq:psieq1}
\end{equation}
which is manipulated to
\begin{equation}
\psi_{n+1}(z) = \psi_n(z) + z \psi_{n-1}(z)
\label{eq:psieq2}
\end{equation}
Now, let
\begin{eqnarray}
\psi_n(z)  & =  & (-z)^{n/2} \phi_n(z) \nonumber \\
& = & \left(\frac{1}{ 2 \cosh k} \right)^n\phi_n(z)
\label{eq:phidef}
\end{eqnarray}
where in the last line of (\ref{eq:phidef}) use has been made of (\ref{eq:param1}). Substituting from (\ref{eq:phidef}) into (\ref{eq:psieq2}), we end up with the recursion relation
\begin{equation}
\phi_{n+1}(z) - 2 \cosh k \phi_n(z) + \phi_{n-1}(z) =0
\label{eq:phieq1}
\end{equation}
The solution to this equation is
\begin{equation}
\phi_n(z) = A e^{kn} + Be^{-kn}
\label{eq:phisol1}
\end{equation}
In the case that $k$ is imaginary, corresponding to $z<-1/4$, the solutions are complex exponential---or, replacing $k$ by $i k$,
\begin{equation}
\phi_n(z) = A \cos k n + B \sin k n
\label{eq:phisol2}
\end{equation}

\subsection{Boundary conditions}

We require that the density be zero at the two ends of the interval. We set the leftmost end at $n=0$ and the rightmost one at $n=L+1$. Making use of (\ref{eq:hdef}) with $h_n$ expressed in terms of $g_n$, we have
\begin{equation}
\rho_n = (1-g_n)(1-\rho_{n+1})
\label{eq:newrhoeq}
\end{equation}
We ensure that $\rho_0=0$ by requiring $g_0=1$. As for $\rho_{L+1}$, we simply set it equal to zero. The pole in the solution for the density will be due to a pole in $g_L$, which will, according to (\ref{eq:psidef}) follow from $\psi_L(z)$ passing through zero. The boundary condition at $n=0$ will be obtained if we set
\begin{equation}
\psi_0(z) = \psi_1(z) =1
\label{eq:bc0}
\end{equation}
Making use of the above equation and Eq.  \eqref{eq:psidef} we see $g_0=0$
The pole at $n=L$ follows from the requirement that
\begin{equation}
\psi_L(z)= \phi_L(z) =0
\label{eq:bcL}
\end{equation}

Given that $k$ will be small in most of the cases of interest to us, our solution  in the uniform case will be generated from
\begin{equation}
\phi_n = \sin (l \pi (n+1) /(L+1))
\label{eq:phisol3}
\end{equation}
with $l$ an integer. Here is how the argument goes. We start with $\phi_n$ of the form $\sin (kn + \theta)$. Then, given that the ratio $\psi_1/\psi_0$ is equal to one, making use of the relationship (\ref{eq:phidef}) between $\psi_n$ and $\phi_n$, we have
\begin{equation}
\frac{\sin (k+ \theta)}{2 \cos k \sin \theta} =1
\label{eq:atzero1}
\end{equation}
Anticipating that both $k$ and $\theta$ will be small, we expand the sine and cosine functions in terms of their arguments, and we end up with
\begin{equation}
\frac{k+ \theta}{2 \theta} =1
\label{eq:atzero2}
\end{equation}
The solution to this equation is $\theta=k$. Requiring that $kL + \theta = l  \pi$, we end up with
\begin{equation}
k=  l \pi/(L+1)
\label{eq:kvals}
\end{equation}
Substituting the results into the argument of the sine function, we end up with (\ref{eq:phisol3}).

\subsection{Reconstruction of the density: general formulas}
\label{page:recb}
The next step will be to reconstruct the density from the function $g_n(z)$. We can do this formally, by noting that the general form of the equation (\ref{eq:rhoeq1}) is
\begin{equation}
\rho_{n+1} = \alpha_n + \beta_n \rho_{n}
\label{eq:genrho1}
\end{equation}
We can solve this equation by analogy with the solution of a linear differential equation. We start by introducing a new variable
\begin{equation}
R_n = \rho_n \prod_{m=m_0}^{n-1} \beta_m^{-1}
\label{eq:genrho2}
\end{equation}
Making use of this new variable, (\ref{eq:genrho1}) is recast into the recursion relation
\begin{eqnarray}
R_{n+1}  &= &  R_n + \alpha_n \prod_{m=m_0}^{n} \beta_m^{-1} \nonumber \\
& \equiv & R_n + \gamma_n
\label{eq:genrho3}
\end{eqnarray}
The solution of this very simple recursion relation is
\begin{equation}
R_n = \sum_{l=l_0}^{n-1} \gamma_l
\label{eq:genrho4}
\end{equation}
In the case of the product in (\ref{eq:genrho2}), the understanding is that if $n-1<m_0$, then the product is replaced by $\prod_{m=n-1}^{m_0} \beta_m$ and similarly for the product in (\ref{eq:genrho3}). As for the sum in (\ref{eq:genrho4}), if $n-1<l_0$ then the sum is replaced by $-\sum_{l=n}^{l_0} \gamma_l$. Finally, if $n-1=m_0$ in (\ref{eq:genrho2}), then the product is replaced by $\beta_{n-1=m_0}$, and similarly for the product in (\ref{eq:genrho3}). Note that all considerations are simplified if we take $m_0=l_0=0$.

Ultimately, we recover the density $\rho_n$ through the inverse of (\ref{eq:genrho2}), with one proviso. Taking $m_0=l_0=0$, we note that a full solution to (\ref{eq:genrho1}) is of the form
\begin{equation}
\rho_n = \left( C + \sum _{l=0}^{n-1} \gamma_l\right) \prod_{m=0}^{n-1}  \beta_m
\label{eq:genrho5}
\end{equation}
with $C$ an arbitrary constant. In the case of interest, the equation corresponding to (\ref{eq:genrho1}), as given by (\ref{eq:rhoeq1}), is
\begin{equation}
\rho_{n+1} = 1 - h_n^{-1} \rho_n
\label{eq:particularrho1}
\end{equation}
so here,
\begin{eqnarray}
\alpha_n & = & 1 \label{eq:particularrho2} \\
\beta_n & = & - h_n^{-1} \label{eq:particularrho3}
\end{eqnarray}

\subsection{Specific relations for the reconstruction of the density}

Given the relationships (\ref{eq:psidef}) and (\ref{eq:phidef}), we can write
\begin{equation}
1+zg_n(z) = \frac{1}{2 \cosh k} \frac{\phi_{n+1}(z)}{\phi_n(z)}
\label{eq:spec1}
\end{equation}
Solving for $g_n(z)$, we have
\begin{equation}
g_n(z) = - 4 \cosh^2 k \left(\frac{1}{2 \cosh k}\frac{\phi_{n+1}(z)}{\phi_n(z)} -1 \right)
\label{eq:spec2}
\end{equation}
Now, we use (\ref{eq:phisol3}), with $k$ general and assume trigonometric  functions ($k \rightarrow ik$).Then,
\begin{equation}
\phi_n(z) = \sin k(n+1)
\label{eq:spec3}
\end{equation}
Where, recall, $z=-1/4 \cos^2k$. Substituting this into (\ref{eq:spec2}), we have
\begin{eqnarray}
h_n(z) & = &1-g_n(z) \nonumber \\ & = & 1 - 4 \cos^2 k + 2 \cos k \frac{\sin k(n+2)}{\sin k(n+1)} \nonumber \\
& = &  - \frac{\sin k(n-1)}{\sin k(n+1)}
\label{eq:spec4}
\end{eqnarray}

As our next step, we will implement the recursion relation repeatedly. Starting with a particular value of $n$, which we will set equal to $L$, and assuming $\rho_{L+1}=0$, we have
\begin{eqnarray}
\rho_{L} &=& \frac{\sin k(L-1)}{\sin k(L+1)} (\rho_{L+1} -1) \nonumber \\
& = & -\frac{\sin k(L-1)}{\sin k(L+1)}
\label{eq:specrec1}
\end{eqnarray}
If we apply this recursion relation down to $\rho_{L-3}$, we end up with
\begin{eqnarray}
\lefteqn{\rho_{L-3}} \nonumber \\ & = &  -\frac{\sin k(L-4)}{\sin k(L-2)} - \frac{\sin k(L-4) \sin k(L-3)}{\sin k(L-2) \sin k(L-1)} - \nonumber \\ && \frac{\sin k(L-4) \sin k(L-3)}{\sin k(L-1) \sin k(L)} - \frac{\sin k(L-4) \sin k(L-3)}{\sin k(L) \sin k(L+1)} \nonumber \\
\label{eq:specrec2}
\end{eqnarray}
There will clearly be a pole in the right hand side of (\ref{eq:specrec2}) when $\sin k(L+1) =0$. However, the denominators in other terms that expression also pass through zero, at different values of $k$. As we will see in our discussion of the more general case, those denominators do not give rise to additional poles.

\subsection{The general case}

We now turn to the way in which the reconstruction of the density works in the most general case. That is, we look into what happens when the fugacity varies from position to position. In that case, the recursion relation can be written in the form
\begin{equation}
g_{n+1} = \frac{1}{1+z_n g_n}
\label{eq:gencase1}
\end{equation}
If we write
\begin{equation}
1+z_n g_n = \frac{\psi_{n+1}}{\psi_n}
\label{eq:gencase2}
\end{equation}
we obtain the following relationship:
\begin{equation}
\psi_{n+2} = \psi_{n+1} + z_{n+1} \psi_n
\label{eq:gencase3}
\end{equation}
Then,
\begin{eqnarray}
h_n & = & 1- g_n \nonumber \\
& = & z_{n-1}\frac{\psi_{n-2}}{\psi_n}
\label{eq:gencase4}
\end{eqnarray}
We can reconstruct the density in the same way as we did in the case of no potential. Carrying out the same procedure as above, we find
\begin{eqnarray}
\lefteqn{\rho_{L-3}} \nonumber \\
& = & z_{L-4}\frac{\psi_{L-5}}{\psi_{L-3}} -z_{L-4} z_{L-3}\frac{\psi_{L-5} \psi_{L-4}}{\psi_{L-3} \psi_{L-2}}  \nonumber \\ && +z_{L-4} z_{L-3}z_{L-2}\frac{\psi_{L-5} \psi_{L-4}}{\psi_{L-2} \psi_{L-1}} \nonumber \\
&& - z_{L-4} z_{L-3}z_{L-2}z_{L-1}\frac{\psi_{L-5} \psi_{L-4}}{\psi_{L-1} \psi_{L}}
\label{eq:gencase5}
\end{eqnarray}
Once again, the pole in the density results from a zero in $\psi_L$. To see that no other zeros lead to a pole, we consider the case of $\psi_{L-2}$. The terms in (\ref{eq:gencase5}) in which that function appears combine as follows.
\begin{eqnarray}
\lefteqn{-z_{L-4}z_{L-3} \psi_{L-5} \psi_{L-4}\frac{1}{\psi_{L-2}} \left(\frac{1}{\psi_{L-3}} - \frac{z_{L-2}}{\psi_{L-1}} \right)} \nonumber \\
& = & -z_{L-4}z_{L-3} \psi_{L-5} \psi_{L-4}\frac{1}{\psi_{L-2}} \left(\frac{\psi_{L-1}- z_{L-2} \psi_{L-3}}{\psi_{L-1}\psi_{L-3}} \right) \nonumber \\
& = & -z_{L-4}z_{L-3} \psi_{L-5} \psi_{L-4}\frac{1}{\psi_{L-2}} \frac{\psi_{L-2}}{ \psi_{L-1} \psi_{L-3}}
\label{eq:gencase6}
\end{eqnarray}
where the last line of (\ref{eq:gencase6}) follows from (\ref{eq:gencase3}) with a suitable adjustment of $n$.

We can alter the equations for $\psi_m$ and $\rho_m$ by introducing a modification of the function $\phi_n$. Our new and generalized version is via the following alteration of (\ref{eq:phidef})
\begin{equation}
\psi_n=  \phi_n \prod_{m=1}^n (-z_n)^{1/2}
\label{eq:gencase7}
\end{equation}
Then, (\ref{eq:gencase3}) becomes
\begin{equation}
(-z_{n+2})^{1/2} \phi_{n+2} = \phi_{n+1} - (-z_{n+1})^{1/2} \phi_n
\label{eq:gencase8}
\end{equation}
Furthermore, the last term in (\ref{eq:gencase6}), the term in which there is a pole resulting from a zero of $\psi_L$---and hence $\phi_L$---reduces to
\begin{equation}
-\frac{(-z_{L-4})^{1/2}}{(-z_L)^{1/2}} \frac{\phi_{L-5} \phi_{L-4}}{\phi_{L-1} \phi_L}
\label{eq:gencase9}
\end{equation}
\label{page:rece}

\section{Relation to the Schr\"{o}dinger equation} \label{sec:sch}

The recursion relation (\ref{eq:phieq1}), and particularly the solutions (\ref{eq:phisol1}) and (\ref{eq:phisol2}), strongly suggest a relationship between the equations that we solve for the quantity $g_n(z)$ and thence the density $\rho_n(z)$ and the Schr\"{o}dinger equation for a free particle. In fact, when the variable $k$ is small and imaginary, Eq. (\ref{eq:phieq1}) becomes
\begin{equation}
\phi_{n+1}(z) - 2 \phi_n(z)(1-k^2/2) + \phi_{n-1}(z) =0
\label{eq:sch1}
\end{equation}
which reduces, under the assumption of slowly-varying $\phi_n(z)$, to
\begin{equation}
- \frac{d^2 \phi_n(z)}{dn^2} = k^2 \phi_n(z)
\label{eq:sch2}
\end{equation}

Now, consider (\ref{eq:gencase8}). We write
\begin{equation}
z_n=-\frac{1}{4 \cos^2 k}e^{- \beta u_n}
\label{eq:sch3}
\end{equation}
Expanding in $k$ and assuming that both $\phi_n(z)$ and $u_n$ are slowly varying functions of $n$, this recursion relation becomes
\begin{equation}
-\frac{d^2 \phi_n(z)}{dn^2} + \left(e^{\beta u_n}-1 \right) \phi_n(z) = k^2 \phi_n(z)
\label{eq:sch4}
\end{equation}
If $ \beta u_n$ is small, the second term on the left hand side of (\ref{eq:sch4}) is just the standard potential energy  contribution to the Schr\"{o}dinger equation---with the multiplicative factor $\beta$. Under the conditions described above, the recursion relations and, more particularly, the density through (\ref{eq:gencase9}), are obtained via the solution to the Schr\"{o}dinger equation. In fact, again, if $\beta u_n$ is small, the residue of the pole in the density as a function of $z$ is given by
\begin{equation}
-\phi_n(z)^2 \mathop{\rm Res}(1/ \phi_{L-1}(z) \phi_L(z))
\label{eq:sch5}
\end{equation}
In fact, the residue function can be directly related to the normalization of the solutions to the equation (\ref{eq:sch4}). This fact can be established by appealing to a standard result for the normalization of the solutions to the one-dimensional Schr\"{o}dinger equation \cite{messiah}. This argument is contained in Appendix \ref{app:norm1}. For an outline of a demonstration based on the discrete equations, see Appendix \ref{app:normalizationdis}.

\section{Alternate derivation of the Schr\"{o}dinger equation formalism from a gradient expansion} \label{sec:gradexpand}

The fact that we are led by the developments in Sections \ref{sec:newmethod} and \ref{sec:sch} from a discrete version of the one-dimensional lattice gas problem as formulated by Percus and Vanderlick to the continuous Schr\"{o}dinger equation suggests the possibility of a direct route from the constitutive equations of the continuous hard rod gas to the Schr\"{o}dinger equation approach. In fact, guided by what has been described previously in this article, one can outline just such a development, based on a gradient expansion of (\ref{eq:van1:nh}) and (\ref{eq:van1:hh}). We start with an analysis of the second of those two equations
\begin{eqnarray}
h(x)&=&e^{\beta(\mu-u(x))} \exp \left[-\int_{x-\sigma}^{x}{h(t)dt} \right] \nonumber \\
& \equiv & z(x)\exp \left[-\int_{x-\sigma}^{x}{h(t)dt} \right]
\label{eq:gradappge1}
\end{eqnarray}
where
\begin{equation}
z(x) = z e^{- \beta u(x)}
\label{eq:gradappzdef}
\end{equation}
When the system is uniform, the singularity occurs on the vicinity of $z(x) = z_0 = -1/ \sigma e$, $h(x) =h_0 =-1/\sigma$. We rewrite the (\ref{eq:gradappge1}) as follows
\begin{eqnarray}
\lefteqn{\sigma h(x) e^{- \sigma h(x)}} \nonumber \\ & = & \sigma (-1/ \sigma + \delta h(x)) e^{\sigma (-1/ \sigma + \delta h(x))} \nonumber \\
& = & -(1- \sigma \delta h(x)) e^{-1} e^{\sigma  \delta h(x)} \nonumber \\
& = & -e^{-1} (1- \sigma \delta h(x) )(1 + \sigma \delta h(x) +  \frac{1}{2} \sigma^2 \delta h(x)^2 + \cdots ) \nonumber \\
& = & -e^{-1} (1+ \frac{1}{2} \sigma^2 \delta h(x)^2 + \cdots) \nonumber \\
& = & \sigma (-1/ \sigma e + \delta z) e^{- \beta u(x)}\exp \left[ - \int_{x- \sigma}^{x} h(t) dt - \sigma h(x) \right] \nonumber \\
& = & -e^{-1}(1- \sigma e \delta z)e^{- \beta u(x)} e^{\sigma^2 d \delta h(x)/dx/2 + \cdots} \nonumber \\
& = & -e^{-1}(1- \sigma e \delta z)(1 - \beta u(x) + \cdots ) \nonumber \\ && \times (1+ \frac{1}{2} \sigma^2 \frac{d \delta h(x)}{dx} + \cdots)
\label{eq:gradappge2}
\end{eqnarray}
Equating the fifth and eighth line of (\ref{eq:gradappge2}), and expanding to second order in $\sigma$ and first order in $\beta u(x)$, we end up with the following equation
\begin{equation}
1 - \frac{1}{2} \sigma^2 \delta h(x)^2 = 1 - \sigma e \delta z + \frac{1}{2} \sigma^2 \frac{d \delta h(x)}{dx} - \beta u(x)
\label{eq:gradappge3}
\end{equation}
or
\begin{equation}
- \frac{d \delta h(x)}{dx} - \delta h(x)^2 + \frac{2}{\sigma^2} \beta u(x) = - \frac{2e}{\sigma} \delta z
\label{eq:gradappge4}
\end{equation}

If we set
\begin{equation}
\delta h(x) = \frac{1}{2} W(x)
\label{eq:gradappWdef}
\end{equation}
then (\ref{eq:gradappge4}) becomes
\begin{equation}
- \frac{dW(x)}{dx} - \frac{1}{2} W(x)^2 + \frac{4}{\sigma^2} \beta u(x) = - \frac{4e}{\sigma} \delta z
\label{eq:gradappge5}
\end{equation}
Let
\begin{eqnarray}
W(x) &=& 2\frac{\psi^{\prime}(x)}{\psi(x)} \nonumber \\
& = & \frac{d}{dx} \ln \left[ \psi(x)^2 \right]
\label{eq:gradapppsidef}
\end{eqnarray}
Then,
\begin{equation}
\frac{dW(x)}{dx} + \frac{1}{2} W(x)^2 = 2 \frac{d^2 \psi(x)/dx^2}{\psi(x)}
\label{eq:gradappWderiv}
\end{equation}
and (\ref{eq:gradappge5}) becomes
\begin{equation}
- \frac{d^2 \psi(x)}{dx^2} + \frac{2}{\sigma^2} \beta u(x) \psi(x) = - \frac{2e}{\sigma}  \delta z  \ \psi(x)
\label{eq:gradappge6}
\end{equation}

We next turn to the other constitutive equation, (\ref{eq:van1:nh}). To analyze this relation, we multiply both sides by $\sigma$ and perform the same kind of expansion as we did on (\ref{eq:gradappge1}). To facilitate this expansion, we replace $\rho(x)$ by $- \rho(x)$, anticipating that the density becomes negative in the regime of interest. In fact, it will become large and negative, as we are interested in the behavior in the vicinity of a pole. Performing the gradient expansion utilized above, (\ref{eq:van1:nh}) becomes
\begin{eqnarray}
\lefteqn{\sigma ( - 1/\sigma + \delta h(x)) (1 +\sigma \rho(x) + \frac{1}{2}\sigma^2 \frac{d\rho(x)}{dx} + \cdots) } \nonumber \\ & = & -(1- \sigma \delta h(x) )(1+ \sigma \rho(x) + \frac{1}{2} \sigma^2 \frac{d\rho(x)}{dx} + \cdots) \nonumber \\
& = & -(1 + \sigma \rho(x) - \sigma^2 \rho(x) \ \delta h(x) + \frac{1}{2} \sigma^2 \frac{d \rho(x)}{dx}  \nonumber \\ &&- \sigma \delta h(x) + \cdots) \nonumber \\
&=& - \sigma \rho(x)
\label{eq:gradappge8}
\end{eqnarray}
The neglected terms in (\ref{eq:gradappge8}) are higher order in $\sigma$. Equating the last line in (\ref{eq:gradappge8}) to the next-to-last line, we end up with the equation
\begin{equation}
\frac{1}{2} \sigma^2 \frac{d \rho(x)}{dx} - \sigma^2 \rho(x) \delta h(x) = \sigma \delta h(x) -1
\label{eq:gradappge9}
\end{equation}
Making use of (\ref{eq:gradappWdef}), this equation becomes
\begin{equation}
\frac{d \rho(x)}{dx} - W(x) \rho(x) = \frac{2}{\sigma^2} (\sigma \delta h(x) -1)
\label{eq:gradappge10}
\end{equation}
Note that the solution to the homogeneous version of (\ref{eq:gradappge10}) is
\begin{eqnarray}
\rho(x) & = & \exp \left[\int^x_{x_0} W(x^{\prime}) dx^{\prime}\right]\nonumber \\
& = &  \exp\left[\ln (\psi(x)^2  - \psi(x_0)^2 )\right] \nonumber \nonumber \\
& = & A \psi(x)^2
\label{eq:gradappge11}
\end{eqnarray}
where we have made use of (\ref{eq:gradapppsidef}).

To further analyze (\ref{eq:gradappge10}), we replace $\delta h(x)$ on the right hand side by $\psi^{\prime}(x)/\psi(x)$, as mandated by (\ref{eq:gradappWdef}) and (\ref{eq:gradapppsidef}). If we further set
\begin{equation}
\rho(x) = \psi(x)^2 r(x)
\label{eq:gradappnewrho}
\end{equation}
Eq. (\ref{eq:gradappge10}) becomes
\begin{equation}
\psi(x)^2 \frac{dr(x)}{dx} = \frac{2}{\sigma^2}\left(\sigma \frac{d \psi(x)/dx}{\psi(x)} -1 \right)
\label{eq:gradappge12}
\end{equation}
In keeping with our gradient expansion approach, we focus our attention on second term in parentheses on the right hand side of (\ref{eq:gradappge12}). The equation that results from ignoring the first term is
\begin{equation}
r(x)  = - \frac{2}{\sigma^2} \int_{x_0}^{x}\frac{dx^{\prime}}{\psi(x^{\prime})^2}
\label{eq:gradappge13}
\end{equation}
and we obtain the following expression for $\rho(x)$
\begin{equation}
\rho(x) = -\frac{2}{\sigma^2}\psi(x)^2  \int_{x_0}^{x}\frac{dx^{\prime}}{\psi(x^{\prime})^2}
\label{eq:rouz1}
\end{equation}
We now note that the function
\begin{equation}
\phi(x) =  \psi(x) \int_{x_0}^{x} \frac{dx^{\prime}}{\psi(x^{\prime})^2}
\label{eq:rouz2}
\end{equation}
is also  a solution to Eq. (\ref{eq:gradappge6}), and that it has a Wronskian of one with the function $\psi(x)$, in that $\phi^{\prime}(x) \psi(x) - \psi^{\prime}(x) \phi(x) =1$ \cite{polyanin}. Both of these properties can be verified by direct substitution. Note that they are independent of the lower bound of integration, $x_0$. A general version of (\ref{eq:rouz2}) is
\begin{equation}
\rho(x) = - \frac{2}{\sigma} \psi(x) ( \phi(x) + A \psi(x))
\label{eq:rouz4}
\end{equation}
We will assume that $\psi(x)$ satisfies the boundary condition $\psi(0)=0$, which yields a $\rho(x) $ that is also zero at that interval boundary. To insure that $\rho(L)=0$, we set the constant $A$ equal to $- \phi(L)/ \psi(L)$. There is then a pole in the density when $\psi(L)=0$. To determine the residue at that pole, we note the following
\begin{eqnarray}
\frac{2}{\sigma^2} \psi(x)^2 \frac{\phi(L)}{\psi(L)} & = & \frac{2}{\sigma^2} \psi(x)^2 \frac{ \phi(L)}{\psi(L)} \frac{\psi^{\prime}(L)}{\psi^{\prime}(L)} \nonumber \\ & = & \frac{2}{\sigma^2} \psi(x)^2 \frac{\phi^{\prime}(L) \psi(L) -1}{\psi(L) \psi^{\prime}(L)}
\label{eq:rouz5}
\end{eqnarray}
The last line of (\ref{eq:rouz5}) follows from the Wronskian relation between $\phi(x)$ and $\psi(x)$. The pole term on the right hand side is the one going as $(\psi(L) \psi^{\prime}(L))^{-1}$. Following the reasoning in Appendix \ref{app:norm1} (see especially (\ref{eq:pp4})) we find that this generates the factor $- \sigma/(2e) 1/((\delta z - \delta z_0) \int_0^L \psi(x)^2 dx)$, where $\delta z_0$ is the value (actually, one of the values) of $\delta z$ at which $\psi(L)=0$. Thus, the residues at the poles incorporate the normalization of the eigenfunctions there, as was the case in the discrete version of the model.

\section{Application of the Schr\"{o}dinger equation fomalism: the case of a delta-function potential} \label{sec:delfun}
\label{sec:delfun1}

In a case of  particular interest to us the potential energy, $u_n$, is non-zero at one site, $n_0$. Here, it is not correct to treat the potential energy as either always small or slowly-varying.  However, as we will see, the connection between solutions to the Schr\"{o}dinger equation and the density as constructed from discrete recursion relation still holds. The recursion relation to which this potential energy leads is
\begin{eqnarray}
g_{n_0+1}  &= &  \frac{1}{1+ze^{-\beta u} g_{n_0}} \nonumber \\
& \equiv & \frac{1}{1+z^{\prime} g_{n_0}}
\label{eq:pot1}
\end{eqnarray}
Rewriting the $g$'s in terms of $\phi$'s, we end up with the relation
\begin{eqnarray}
\lefteqn{\phi_{n_0+2} } \nonumber \\
& = & 2 \cos k \phi_{n_0+1} - \phi_{n_0}  \nonumber \\ &&+ \phi_{n_0} \left(1- \frac{z^{\prime}}{z} \right)\frac{2 \cos k \phi_{n_0} - \phi_{n_0+1}}{2 \cos k (1-z^{\prime}/z) \phi_{n_0} + \phi_{n_0+1} z^{\prime}/z} \nonumber \\
& = & \phi^0_{n_0+2} \nonumber \\ && + \phi_{n_0}\left(1-\frac{z^{\prime}}{z} \right)\frac{\phi_{n_0-1}}{2 \cos k(1-z^{\prime}/z) \phi_{n_0}+ \phi_{n_0+1}z^{\prime}/z} \nonumber \\
\label{eq:deltaeq1}
\end{eqnarray}
The quantity $\phi^0_{n_0+2}$ is the value of $\phi_{n_0+2}$ in the absence of the delta function potential. If that potential is small, then $(1-z^{\prime}/z)$ will be small, and the right hand side of (\ref{eq:deltaeq1}) becomes
\begin{equation}
\phi^0_{n_0+2} + \phi_{n_0}\left( \frac{z}{z^{\prime}}-1\right) \phi_{n_0-1}/\phi_{n_0+1}
\label{eq:deltaeq2}
\end{equation}

Now, we once again take the ratio with which we started. This leads us to the equation
\begin{equation}
\frac{\phi_{n_0+2}}{\phi_{n_0+1}} = \frac{\phi^0_{n_0+2}}{\phi_{n_0+1}}+\left(\frac{z}{z^{\prime}}-1 \right) \frac{\phi_{n_0} \phi_{n_0-1}}{\phi_{n_0+1}^2}
\label{eq:deltaeq3}
\end{equation}
We will consider two cases: extended states and the possibility of a bound state. The latter solution exists if the potential is attractive and exceeds a threshold value.

\subsubsection{Extended state solutions}

The assumption that we now work with is that the solutions to the equations have the following form
\begin{equation}
\phi_n= \left\{ \begin{array}{ll} a \sin(k(n+1)) & n\le n_0+1 \\ b \sin(k(n+1) + \theta) & n \ge n_0+1 \end{array} \right.
\label{eq:phiform1}
\end{equation}
Note that there are two possible forms for of $\phi_n$ at $n=n_0+1$. That is, $a \sin(k(n_0+2))= b \sin(k(n_0+2) + \theta)$. The quantity $\phi^0_{n_0+2}$ is, according to (\ref{eq:phiform1}), equal to $a \sin(k(n+3))$. Given all this, (\ref{eq:deltaeq3}) becomes
\begin{eqnarray}
\lefteqn{\frac{\sin(k(n_0+3) + \theta)}{\sin(k(n_0+2) + \theta)}} \nonumber \\ & =&  \frac{\sin(k(n_0+3))}{\sin(k(n_0+2))} + \left(\frac{z}{z^{\prime}}-1 \right) \frac{\sin(k(n_0 + 1)) \sin(kn_0)}{\sin(k(n_0+2))^2} \nonumber \\
\label{eq:deltaeq4}
\end{eqnarray}
We now consider the most general version of (\ref{eq:deltaeq4}) as a condition on the phase shift $\theta$. That is, we look at the equation
\begin{equation}
\frac{\sin(k(n_0+3) + \theta)}{\sin(k(n_0+2) + \theta)} = A
\label{eq:geneq1}
\end{equation}
Expanding the sine functions and solving for $\theta$, we end up with the following result:
\begin{equation}
\tan \theta = \frac{\sin (k(n_0+3)) - A \sin (k(n_0+2))}{A \cos (k(n_0+2)) - \cos (k(n_0+3))}
\label{eq:geneq2}
\end{equation}
Inserting the right hand side of (\ref{eq:deltaeq4}) into (\ref{eq:geneq2}) as a substitute for the quantity $A$, we end up with the result for $\tan \theta $:
\begin{widetext}
\begin{eqnarray}
\tan \theta =- \frac{\left(z/z^{\prime}-1 \right) \sin(k(n_0+1)) \sin(k n_0)}{\sin k +(z/z^{\prime}-1) \sin (k(n_0+1)) \sin(kn_0) \cos(k(n_0+2))/\sin(k(n_0+2))}
\label{eq:deltaeq5}
\end{eqnarray}
\end{widetext}

We are now in a position to work out the allowed values of the quantity $k$. Given the boundary condition (\ref{eq:bcL}), the requirement on $k$ is
\begin{eqnarray}
\sin(k(L+1) +  \theta(k)) =0
\label{eq:kreq1}
\end{eqnarray}

\label{page:delfun2}

\subsubsection{The bound state}

Here, we assume hyperbolic functions, corresponding to a form for $k$ that places the fugacity, $z$,  closer to the origin than $-1/4$. The solution to the recursion relations for $\phi_n(z)$ will then be
\begin{equation}
\phi_n(z) = \left\{ \begin{array}{ll}   a \sinh (k(n+1)) & n\leq n_0+1 \\ b e^{-kn} & n \geq n_0+1 \end{array} \right.
\label{eq:bsa1}
\end{equation}
This solution is appropriate to a system in which the length of the region to which the rods are confined is arbitrarily great. Applying (\ref{eq:deltaeq3}) to this conjectured solution, we end up with the relationship
\begin{eqnarray}
\lefteqn{e^{-k}} \nonumber \\ &=& \frac{\sinh  (k(n_0+3))}{\sinh(k(n_0+2))}   \nonumber \\ && +\left(\frac{z}{z^{\prime}}-1 \right) \frac{\sinh(k(n_0+1)) \sinh (kn_0)}{\sin^2(k(n_0+2))}
\label{eq:bsa2}
\end{eqnarray}
which can be manipulated to
\begin{eqnarray}
\lefteqn{\sinh k \sinh(k(n_0+2))} \nonumber \\ & =& \left(1-\frac{z}{z^{\prime}} \right) \sinh(kn_0) \sinh(k(n_0+1)) e^{-k(n_0+2)} \nonumber \\
\label{eq:bsa3}
\end{eqnarray}

\subsubsection{Limiting cases}

Two limits are of interest to us. First, if $n_0 \gg 1$, then (\ref{eq:bsa3}) reduces to
\begin{equation}
1-\frac{z}{z^{\prime}} = 2 \sinh k e^{3k}
\label{eq:bs2}
\end{equation}
Given $1-z/z^{\prime} = 1-e^{\beta u}$, we note that (\ref{eq:bs2}) has a solution of the type desired only if $u<0$. If we expand the left hand side of (\ref{eq:bs2}) in $u$ and, assuming negative $u$,  replace the left hand side of (\ref{eq:bs2}) by $-|\beta u|$, we obtain the following equation for $k$
\begin{equation}
e^{4k}-e^{2k} - |\beta u| =0
\label{eq:bs3}
\end{equation}
The solution to this equation is
\begin{equation}
k = \frac{1}{2} \ln \left[ \frac{1}{2} \left(  1 + \sqrt{1+ 4 |\beta u|}\right)\right]
\label{eq:bs4}
\end{equation}
When $k$ is small---the other limit of interest---the right hand side of (\ref{eq:bs4}) can be expanded, and we find
\begin{equation}
k=|\beta u|
\label{eq:bs4a}
\end{equation}

\subsubsection{The case of small $k$}

Continuing in our investigation of the small-$k$ regime, we note that when $k \ll 1$ we can ignore the difference between $\sinh (kn_0)$ and $\sinh(k(n_0+1))$ and $\sinh(k(n_0+2))$ in (\ref{eq:bsa3}), and we can also replace $\sinh k$ by $k$. Then, the condition on $k$ is
\begin{equation}
k=\left( 1- \frac{z}{z^{\prime}}\right) \sinh(kn_0) e^{-k}
\label{eq:bs5}
\end{equation}
This limit can also be applied to the calculation of the properties of bound states. For example, Eq. (\ref{eq:deltaeq5}) reduces to
\begin{equation}
\tan \theta = \frac{-(z/z^{\prime}-1) \sin^2(kn_0)/k}{1+(z/z^{\prime}-1) \sin( kn_0) \cos(kn_0)/k}
\label{eq:bs6}
\end{equation}

As will be verified in Section \ref{sec:nextcalc}, Eqs. (\ref{eq:bs4a})--(\ref{eq:bs6}) are consistent with the equations at which one arrives in the case of a delta function potential in the corresponding Schr\"{o}dinger equation. This is especially the case if we expand the factor $(1-z/z^{\prime})$ to first order in $\beta u$, which corresponds to the magnitude of the Dirac delta function potential.

\section{Delta-function potential well in the vicinity of a surface} \label{sec:nextcalc}

Henceforth, we will assume that the functions, $\phi_n(z)$ can be rewritten in the form $\phi(x,z)$ where $x$ is now a continuous variable, and that those function can be determined by solving the appropriate version of a Schr\"{o}dinger equation. We then assume an attractive delta function potential in the vicinity of the bounding surface at $x=0$, and we rederive the equations satisfied by the bound state and the phase shift in extended states. We start with the standard ``matching condition''
\begin{equation}
\left(\frac{\phi^{\prime}(x)}{\phi(x)} \right)_{x=x_1- \epsilon} - \left(\frac{\phi^{\prime}(x)}{\phi(x)} \right)_{x=x_1+ \epsilon} = V
\label{eq:sd1}
\end{equation}
where $V$ is the strength of the attractive potential at the point $x_1$, which we use as a simpler substitute for the combination $(z/z^{\prime} -1)$. Note that the dependence of $\phi(x,z)$ on $z$ has been suppressed. This practice will be followed throughout this section.

\subsection{The bound state} \label{subsec:boundstate}
The unnormalized bound state is given by
\begin{equation}
\phi_b(x) = \left\{ \begin{array}{ll}  \sinh \kappa x & x<x_1 \\ \sinh \kappa x_1 e^{- \kappa (x-x_1)} & x> x_1\end{array} \right.
\label{eq:sd2}
\end{equation}
Applying the matching condition (\ref{eq:sd1}) to the solution (\ref{eq:sd2}), we end up with the equation for the quantity $\kappa$
\begin{equation}
\coth \kappa x_1 + 1 = \frac{V}{\kappa}
\label{eq:sd3}
\end{equation}
One readily establishes, by looking at small-$\kappa$ limits, that there is now a minimum value of $V$ required to sustain a bound state. In order for this to be possible, we must have $V > 1/x_1$. An alternate, but equivalent, version of (\ref{eq:sd3}) is
\begin{equation}
\kappa - Ve^{- \kappa x_1} \sinh \kappa x_1  =0
\label{eq:sd4}
\end{equation}
This relation is to be compared with(\ref{eq:bs5}).

The normalization of the bound state is obtained by taking the integral
\begin{equation}
\int_0^{\infty} \phi_b(x)^2 dx = \frac{1}{2 \kappa} \left[ \sinh \kappa x_1e^{\kappa x_1} - \kappa x_1 \right]
\label{eq:sd4a}
\end{equation}

\subsection{Extended states}

Here, the states are of the form
\begin{equation}
\phi(x) = \left\{ \begin{array}{ll}\frac{\sin kx}{\sin kx_1} \sin(kx_1 + \theta(k)) & x<x_1 \\ \sin(kx+ \theta(k)) & x>x_1 \end{array} \right.
\label{eq:sd5}
\end{equation}
Note that the form of the extended eigenfunctions is consistent with the normalization considerations laid out in Appendix \ref{app:normalization}.
The equation satisfied by $\theta(k)$ is
\begin{equation}
k \cot kx_1 - k \cot(kx_1 + \theta(k)) = V
\label{eq:sd6}
\end{equation}
After manipulations like those in Section \ref{sec:delfun}, we end up with the equation for $\theta(k)$
\begin{equation}
\tan \theta(k) = \frac{\frac{V}{k} \sin^2 kx_1}{1-\frac{V}{k} \cos kx_1 \sin kx_1}
\label{eq:sd7}
\end{equation}
This relation effectively replicates (\ref{eq:bs6}).

\subsection{The calculation of the generating function }

\begin{figure}[htbp]
\begin{center}
\includegraphics[width=3in]{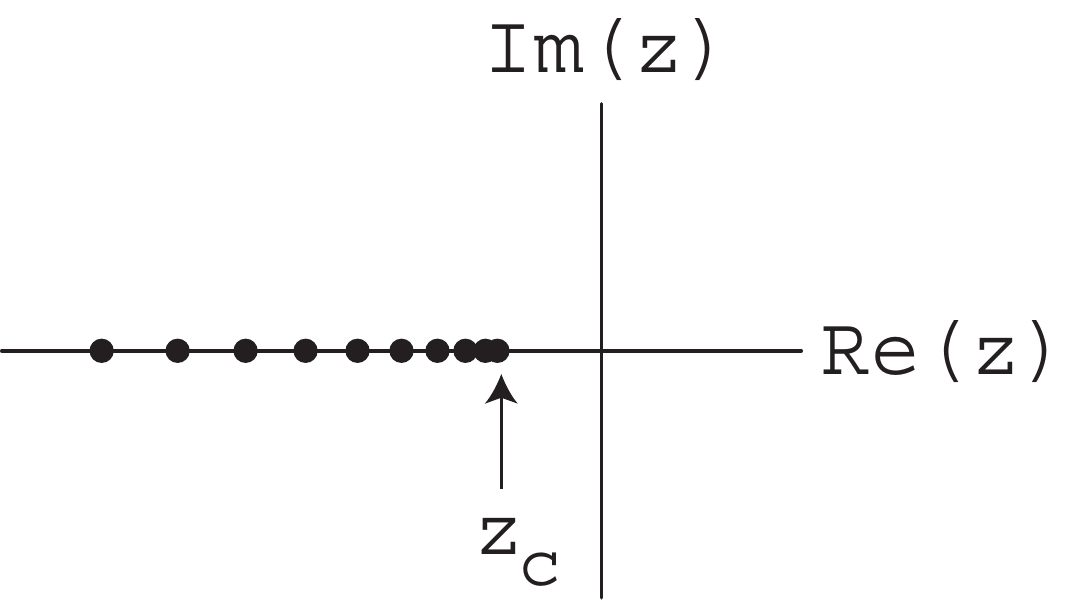}
\caption{The locations of the poles in the density as a function of the fugacity, $z$. The location of $z_c=-1/4$, is indicated.}
\label{fig:polegraph}
\end{center}
\end{figure}

We start by noting that the following holds in the vicinity of the critical point.
\begin{eqnarray}
z  &= & - \frac{1}{4} \frac{1}{\cos^2 k} \nonumber \\
& \rightarrow & - \frac{1}{4}(1+k^2)
\label{eq:res1a}
\end{eqnarray}
Given the values that $k$ takes in a system with constant potential and large system size, $L$, (see (\ref{eq:kvals})), the poles in the density as a function of fugacity, $z$, will be closely-spaced on the negative $z$-axis as indicated in Fig. \ref{fig:polegraph}.

We will start with the generating function for the gas of rods deep in the interior of a constant potential system. We use the Mittag-Leffler theorem \cite{jeffreys} to reconstruct the density from its poles. In the case at hand, we can ignore the spatial structure of the modes, so that the residues can be assumed to be constant. The relevant expression is
\begin{equation}
\frac{1}{L}\sum_{l}\frac{1}{z-z_l}
\label{eq:ml1}
\end{equation}
where quantities $z_l$ are the poles and $1/L$ encapsulates the normalization of the modes. Given (\ref{eq:res1a}), this sum reduces to
\begin{equation}
\frac{1}{2 \pi |z_c|} \int_{-\infty}^{\infty} \frac{dk}{(z+|z_c|)/|z_c|+k^2} = \frac{1}{\sqrt{|z_c|}}\frac{1}{2 \sqrt{|z_c|+z}}
\label{eq:gfback1}
\end{equation}
As previously, $z_c=-1/4$. Obtaining the coefficient of $z^{N}$ in the expansion of the right hand side of (\ref{eq:gfback1}) is a pretty straightforward exercise. I will utilize a method that is, at least initially, a bit more complicated. We start with the standard contour integral-based expression for the coefficient of $z^N$ in the expansion of a function of $z$. As applied to the function at issue here, it is
\begin{equation}
\frac{1}{2\sqrt{|z_c|}}\frac{1}{2 \pi i}\oint \frac{1}{z^{N+1}} \frac{1}{\sqrt{|z_c|+z}} dz
\label{eq:gfback2}
\end{equation}
where the closed contour encircles the origin. We distort the contour so that it wraps around the branch cut that starts at $z=z_c$ and extends to $- \infty$.  If we write $z=-|z_c|-\zeta$ then the integral will look like this, to within overall multiplicative constants.
\begin{eqnarray}
\lefteqn{\int_{0}^{\infty} \frac{1}{(-|z_c|-\zeta)^{N+1}} \frac{1}{\sqrt{\zeta}} d \zeta } \nonumber \\ &=& (-1)^{N+1} \int_0^{\infty} \frac{1}{(|z_c|+\zeta)^{N+1}} \frac{1}{\sqrt{\zeta}} d \zeta
\label{eq:gfback3}
\end{eqnarray}
Now, we introduce the following identity
\begin{equation}
\frac{1}{(|z_c|+ \zeta)^{N+1}} = \frac{1}{N!} \int_0^{\infty}t^Ne^{-t(z_z + \zeta)}dt
\label{eq:gfback4}
\end{equation}
The double integral we now have to perform is
\begin{equation}
(-1)^{N+1} \frac{1}{N!} \int_0^{\infty} dt \int_{0}^{\infty} d \zeta  t^N \frac{e^{-t(|z_c|+ \zeta)}}{\sqrt{\zeta}}
\label{eq:gfback5}
\end{equation}
We will perform the integral over $\zeta$ first. It is
\begin{eqnarray}
\int_0^{\infty} \frac{e^{- t \zeta}}{\sqrt{\zeta}} d \zeta  &= &2 \int_0^{\infty} e^{-ty^2} dy \nonumber \\
& = & \int_{-\infty}^{\infty} e^{-ty^2}dy \nonumber \\
& = & \sqrt{\frac{\pi}{t}}
\label{eq:gfback6}
\end{eqnarray}
The remaining integral is
\begin{eqnarray}
\lefteqn{(-1)^{N+1} \frac{\sqrt{\pi}}{N!} \int_0^{\infty} t^{N-1/2}e^{-|z_c|t} dt} \nonumber \\ &  = &  (-1)^{N+1} \frac{\sqrt{\pi}}{N!} \int_0^{\infty} \exp \left[ (N-1/2) \ln t - t|z_c|\right] dt \nonumber \\
\label{eq:gfback7}
\end{eqnarray}
When $N \gg 1$, we can evaluate the integral by looking for its maximum. The extremum equation that determines the maximizing value of $t$ is
\begin{equation}
\frac{N-1/2}{t} -|z_c| = 0
\label{eq:gfback8}
\end{equation}
The solution to this equation is
\begin{equation}
t=\frac{N-1/2}{|z_c|}
\label{eq:gfback9}
\end{equation}
Substituting this back into the integrand we find for the result of the integral
\begin{widetext}
\begin{eqnarray}
\lefteqn{(-1)^{N+1} \sqrt{\pi} \frac{1}{N!} \exp \left[(N-1/2) \ln \left( \frac{N-1/2}{|z_c|}\right)  - (N-1/2)\right] } \nonumber \\ & = & (-1)^{N+1} \sqrt{\pi} \frac{1}{N!} \exp \left[ (N-1/2) \ln |z_c| +(N-1/2) \ln N - 1/2 -(N-1/2) \right] \nonumber \\ & = & (-1)^{N+1} \sqrt{\pi} \frac{1}{N!} |z_c|^{-(N-1/2)} \exp \left[(N-1/2) \ln N - N \right] \nonumber \\
& = & (-1)^{N+1}\sqrt{\pi} |z_c|^{-(N-1/2)} \exp \left[ (N-1/2) \ln N - N -(N \ln N - N)\right] \nonumber \\
& = & (-1)^{N+1} \sqrt{\pi} |z_c|^{-(N-1/2)} \exp \left[ -(1/2) \ln N\right] \nonumber \\
& = & (-1)^{N+1} \sqrt{\pi} |z_c|^{-(N-1/2)} N^{-1/2}
\label{eq:gfback10}
\end{eqnarray}
\end{widetext}
In the fourth line of (\ref{eq:gfback10}), Stirling's formula for $N!$ was used.

The principal result, that the coefficient of $z_N$ goes as $|z_c|^{-N}N^{-1/2}$, could have been derived considerably more easily. However, the method can be generalized. For example, consider the case of the generating function near a boundary. Here, we have for the generating function
\begin{eqnarray}
\frac{2}{\pi}\int_0^{\infty} \frac{\sin^2 kx}{|z_c| + z + k^2} dk & = & \frac{1}{\pi} \int_0^{\infty}\frac{1-\cos 2kx}{|z_c|+z +k^2} dk \nonumber \\
& = & \frac{1}{2 \pi} \int_{-\infty}^{\infty} \frac{1-\cos 2kx}{|z_c|+z +k^2} dk \nonumber \\
& = & \frac{1-e^{-2\sqrt{|z_c|+z} \ x}}{2 \sqrt{|z_c|+z}}
\label{eq:gfback11}
\end{eqnarray}
We know how to extract the coefficient of $z^N$ in part of the expression on the last line of (\ref{eq:gfback11}). For the additional part, the double integral corresponding to (\ref{eq:gfback5}) is
\begin{equation}
(-1)^{N+1} \frac{1}{N!} \int_0^{\infty} dt \int_0^{\infty} d \zeta \  t^N\frac{e^{-t(|z_c| + \zeta)} \cos (2 \sqrt{\zeta}  \, x)}{\sqrt{\zeta}}
\label{eq:gfback12}
\end{equation}
Again, we perform the integral over $\zeta$ first, changing integration variables as in (\ref{eq:gfback6}). We end up with the integral
\begin{eqnarray}
2\int_0^{\infty}e^{-ty^2} \cos (2yx) dy & = & \int_{-\infty}^{\infty} e^{-ty^2 + 2iyx} dy \nonumber \\
& = &\sqrt{ \frac{\pi}{t} }e^{-x^2/t}
\label{eq:gfback13}
\end{eqnarray}
The final integration to perform is
\begin{equation}
(-1)^{N+1}\frac{\sqrt{\pi}}{N!} \int_0^{\infty} \exp \left[(N-1/2) \ln t -t|z_c| -x^2/t \right]
\label{eq:gfback14}
\end{equation}
The new extremum equation is
\begin{equation}
\frac{N-1/2}{t} -|z_c| +\frac{x^2}{t^2} =0
\label{eq:gfback15}
\end{equation}
The analysis can be short-circuited if we take into account the following facts: \begin{enumerate}

\item The correction to the solution of the equation due to the last term will be small.
\item The effect of the correction on the first two terms in the exponent in (\ref{eq:gfback14}) will also be very small, as $t$ has already been adjusted so that those terms are at an extremum.
\end{enumerate}
This all means that the result of the integration in (\ref{eq:gfback14}) will to be the same as in (\ref{eq:gfback7}), except that there is the additional term
\begin{equation}
-\frac{x^2}{t} =  - \frac{x^2 |z_c|}{N-1/2}
\label{eq:gfback16}
\end{equation}
Combining this with the term we have already evaluated we have for the coefficient of $z^N$ in the case of the hard-rod gas near an end-wall
\begin{equation}
(-1)^{N+1} \sqrt{\pi}|z_c|^{-(N-1/2)} N^{-1/2} \left(1-e^{-|z_c|x^2/(N-1/2)} \right)
\label{eq:gfback17}
\end{equation}
The difference between $N$ and $N+1/2$ can be neglected in the denominator in the exponent in (\ref{eq:gfback17}).

\subsection{Density in a finite interval}

We can also utilize the Mittag-Leffler method to reconstruct the density in the case of a finite interval. Here, the reconstructed density is, to within an overall multiplicative factor
\begin{eqnarray}
\lefteqn{\rho(n)} \nonumber \\ & = & \frac{1}{L} \sum_{m=1}^{\infty} \left(\frac{1}{4} + \frac{(\pi m)^2}{4L^2} \right)^{-N} \sin^2(\pi m n/L)  \nonumber \\ & = & \frac{4^N}{L} \sum_{m=1}^{\infty}e^{-N \ln(1+\pi^2 m^2/L^2)} \sin^2(\pi mn/L) \nonumber \\
& \rightarrow & \frac{4^N}{L} \sum_{m=1}^{\infty} e^{-\pi^2 m^2 N/L^2} \sin^2( \pi m n/L)
\label{eq:polesum1}
\end{eqnarray}
where the explicit value of $z_c$ is used. There are two different limits to consider, based on the ratio $N/L^2$. If $N \gg L^2$, then the sum is dominated by the first term, and the density is
\begin{equation}
\rho(n) = \frac{4^N}{L} e^{- \pi^2N/L^2} \sin^2(\pi n/L)
\label{eq:polelim1}
\end{equation}
On the other hand, if $N \ll L^2$, then the sum is as given by (\ref{eq:gfback17}).

\begin{figure}[htbp]
\begin{center}
\includegraphics[width=3in]{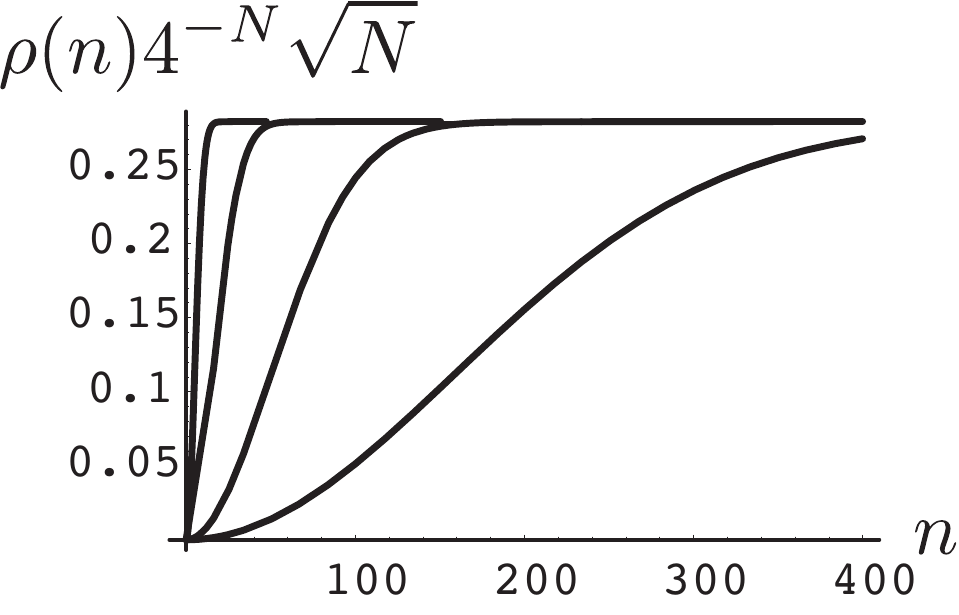}
\caption{The density, as given by the last line of (\ref{eq:polesum1}), as a function of $n$, multiplied by $ 4^{-N} \sqrt{N}$, with $L=10,000$, for the following value of $N$:  50, 500, 5,000, 50,000. The highest curves correspond to the smallest values of $N$. The plot is restricted to the region adjacent to the boundary. }
\label{fig:densplot2}
\end{center}
\end{figure}

Figure \ref{fig:densplot2} shows how the density as given by (\ref{eq:polesum1}) behaves as a function of $n$ when $L=10,000$ for various values of $N$. The function is multiplied by $4^{-N} \sqrt{N}$ so that the various curves tend to  the same value in the interior of the interval when $N$ is small enough.

\begin{figure}[htbp]
\begin{center}
\includegraphics[width=3in]{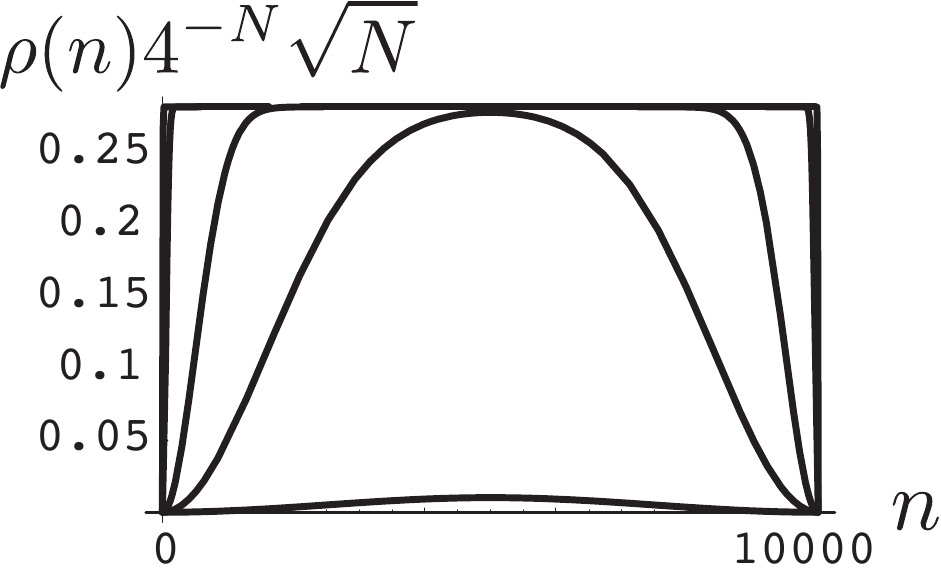}
\caption{The density as given by the last line of (\ref{eq:polesum1}), for a gas of hard rods, normalized as in Fig. \ref{fig:densplot2}. In this case the values of $N$ are 50, 5,000, 500,000, 5,000,000 and 50,000,000. Again, the highest curves correspond to the smallest values of $N$. }
\label{fig:densplot5}
\end{center}
\end{figure}

Figure \ref{fig:densplot5} displays the density profile over the entire interval for a different set of values of $N$.
The dominance of the single Schr\"{o}dinger equation eigenfunction, $\sin(\pi(n+1)/L)$, is evident in $N=50,000,000$ curve.

\subsection{The attractive potential}

We can make use of the results above to perform the integral needed to reconstruct the density as a function of the fugacity. An important precursor to the calculation of the generating function is the reconstruction of the sum $\sum_l \phi_l(x)^2$. The details of this calculation are contained in Appendix \ref{app:sumcalc}. Making use of the results of the results of that appendix, we find that the {\em modification} of the generating function for the density due to the presence of the delta function potential near the boundary is
\begin{eqnarray}
\lefteqn{-\frac{1}{\pi} \int_{-\infty}^{\infty} \sin^2 kx_1 \mathop{\rm Re} \left[ \frac{e^{2ikx}V}{k-V\sin kx_1 e^{ikx_1}}\right] } \nonumber \\ && \times \frac{dk}{(z+|z_c|) + k^2} \nonumber \\ && + \frac{\sin^2 \kappa x_1 e^{-2\kappa (x-x_1)}}{\frac{1}{2 \kappa}\left[ \sinh \kappa x_1 e^{\kappa x_1} - \kappa x_1\right]} \nonumber \\
& = & \frac{\sinh^2 (\sqrt{z+|z_c|} \, x_1)}{\sqrt{z+|z_c|}} \nonumber \\ && \times \frac{e^{-2 \sqrt{z+|z_c|} \, x} V}{\sqrt{z+|z_c|} -  V \sinh(\sqrt{z+|z_c|} \, x_1) e^{-\sqrt{z+|z_c|} \, x_1}} \nonumber \\
\label{eq:gf1}
\end{eqnarray}
In the above equation, the quantity $\kappa$ satisfies (\ref{eq:sd3}) or, equivalently, (\ref{eq:sd4}). This contribution to the generating function is, recall, in addition to the contribution that one derives in the absence of the attractive potential. The results in (\ref{eq:gf1}) are relevant to the case $x>x_1$. Now, the extraction of the actual density at fixed monomer number, $N$, from (\ref{eq:gf1}) entails the kind of contour integral described in Section \ref{sec:dimred}. The result of that integration  depends on the value of $V$. If the potential strength is sufficiently great that there is a bound state, then one can show straightforwardly that there is a pole in the last line of (\ref{eq:gf1}) with a residue that yields the normalized bound state with a prefactor going as $z_*^{-N}$, where
\begin{equation}
\sqrt{|z_c|+z_*}= \kappa
\label{eq:gf2}
\end{equation}
The quantity $\kappa$, again, satisfies the equivalent equations (\ref{eq:sd3}) or, equivalently (\ref{eq:sd4}), for the bound state. Note that the absolute value of $z_*$ is less than $|z_c|$, in that $z_*=\kappa^2-|z_c|$. We are assuming a $\kappa$ that is not too large. If $V$ does not exceed the threshold for a bound state, then things are a bit different. There is no pole in (\ref{eq:gf1}), but rather a branch cut. A detailed calculation, under the assumption $x \sim \sqrt{N}$ and $x_1 \ll \sqrt{N}$, yields
\begin{eqnarray}
\lefteqn{\rho(x)} \nonumber \\ & \propto & |z_c|^{-N}N^{-1/2} \nonumber \\ && \left[1-e^{-|z_c|x^2/N} \left(1-\frac{2V|z_c|x_1^2x/\sqrt{N}}{1-Vx_1} \right)\right]
\label{eq:gf3}
\end{eqnarray}
Note the denominator in the last term in brackets on the left hand side of (\ref{eq:gf3}), a signature of an impending ground state.

\subsection{Attractive potential at the boundaries of a finite interval}

\begin{figure}[htbp]
\begin{center}
\includegraphics[width=3in]{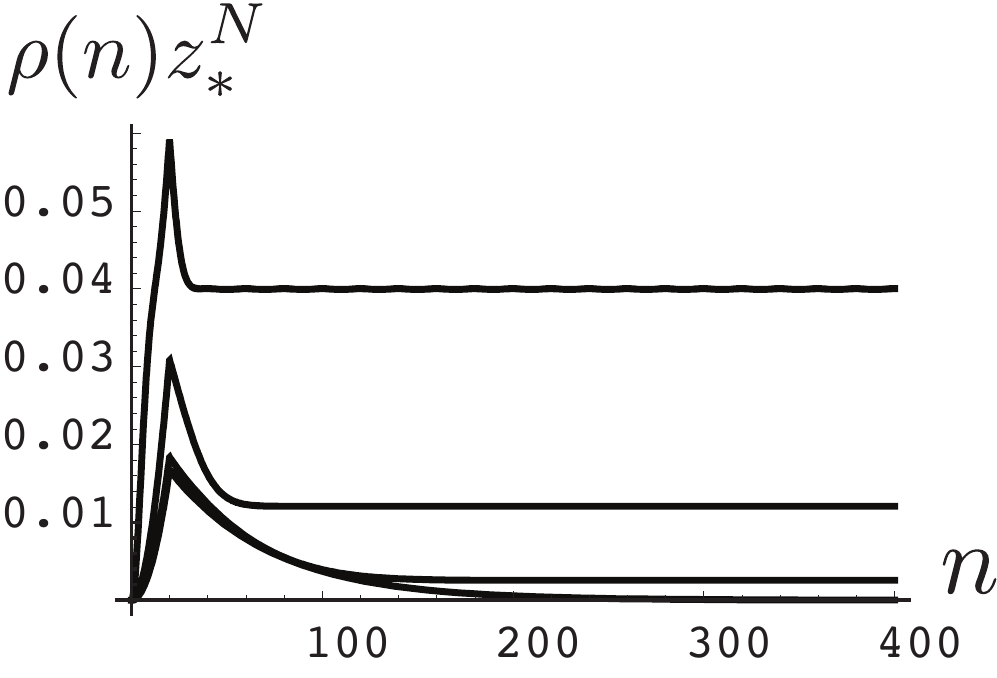}
\caption{The density, $\rho(n)$, for the case of attractive potentials near the bounding surfaces of of a region of length $L=10,000$. The curves correspond to $N=$50, 500, 5,000 and 50,000. The heights of the density curves decrease with increasing $N$. }
\label{fig:attractive1}
\end{center}
\end{figure}

\begin{figure}[htbp]
\begin{center}
\includegraphics[width=3in]{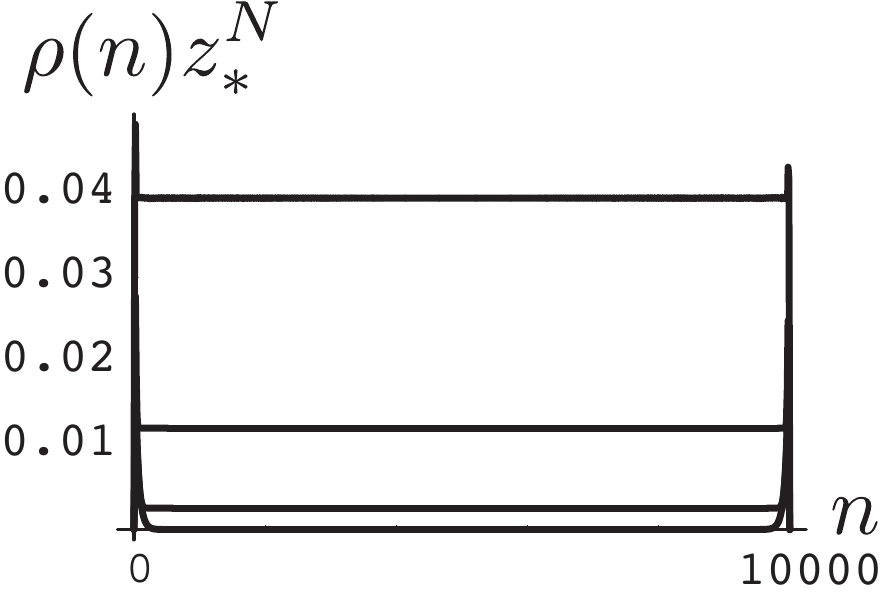}
\caption{The same set of densities graphed in Fig. \ref{fig:attractive1}, this time over the entire interval.}
\label{fig:attractive2}
\end{center}
\end{figure}

Some numerical results serve to illustrate the effects of attractive potentials on the density of branched polymers confined to a finite interval. For example, Fig. \ref{fig:attractive1} shows what the density looks like for various values of the number of monomers, $N$, when there is an attractive potential a distance $n=40$ from the boundaries of a region with extension $L=10,000$. Here, we take $N=$ 50, 500, 5,000 and 50,000.
The figure graphs the density close in to one of the bounding surfaces. In each case, the density is multiplied by $z_*^N$, where $z_*$ is the value of $z$ corresponding to the pole associated with the lowest energy bound state. In the case of Fig. \ref{fig:attractive1}, the attractive potentials are sufficiently strong to ensure two bound states. Figure \ref{fig:attractive2} displays the same set of modified densities, this time over the entire interval.

\begin{figure}[htbp]
\begin{center}
\includegraphics[width=3in]{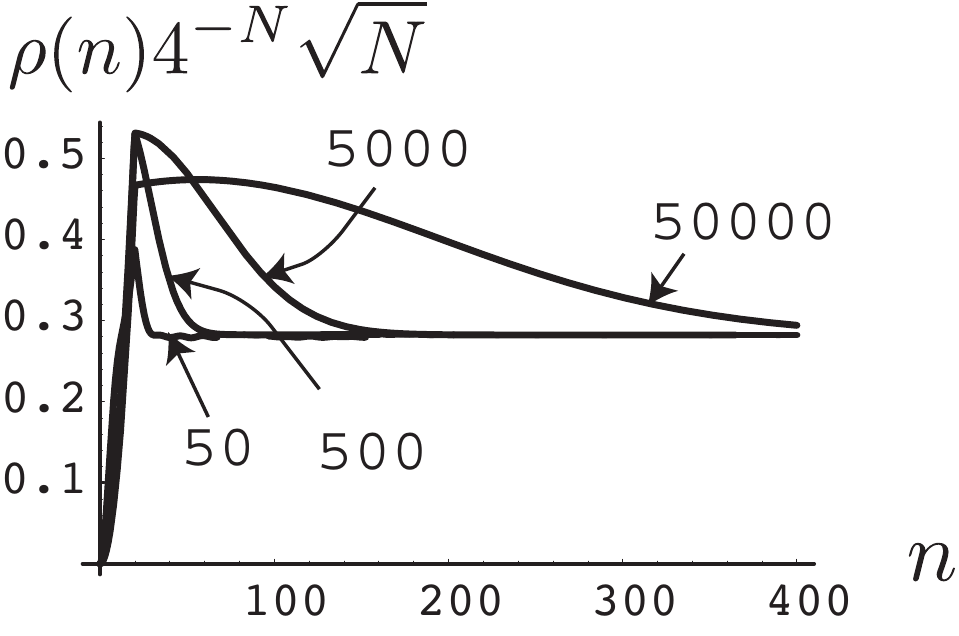}
\caption{The density, normalized as in Fig. \ref{fig:densplot2}, in the vicinity of one of the boundaries of a system with length, $L=10,000$, for  $N$ equal to 50, 500, 5,000 and 50,000. The attractive potential, which is not quite strong enough to induce a bound state, is at a distance 40 from the boundary.}
\label{fig:attractive3}
\end{center}
\end{figure}

Finally, we consider the case of an attractive potential that is not quite strong enough to generate bound states. Here, one might expect to see the influence of a  ``precursor effect.'' Figure \ref{fig:attractive3} graphs the density, normalized as in Fig. \ref{fig:densplot2}. The dependence on number of particles, $N$, which is not as simple and monotonic as in the previous cases, is indicated in the figure.
The dependence of $\rho(n)$ on $N$ in this case is to be contrasted with the graphs of $\rho(n)$ in Fig. \ref{fig:densplot2}, where at $N=5,000$, the normalized density is greatly suppressed with respect to the density at lower values of $N$.

\section{The pressure of rooted trees}

Recall that the generating function for the number of trees with roots at the position $x$ is derivable from the density of a gas of hard-core particles. As we have seen, the form of this density is
\begin{equation}
\sum_k\frac{\psi_k(x)^2}{z_k + z}
\label{eq:dens1}
\end{equation}
The functions $\psi_k(x)$ are solutions to a Schr\"{o}dinger-like equation that are, furthermore, normalized. If we sum over all possible locations of the roots, we end up with a generating function directly derivable from the sum for all rooted trees in the interval. This generating function is
\begin{equation}
\sum_k\frac{1}{z_k+z}
\label{eq:dens2}
\end{equation}
We are interested in the coefficient of $z^N$ in this sum, which leads us to the following analogue of the partition function for rooted trees in the interval:
\begin{equation}
\sum_kz_k^{-N}
\label{eq:dens3}
\end{equation}
There are additional combinatorial factors, but, as we will be looking at large $N$ and taking a log, they turn out to be unimportant.

From the expression above, we obtain the following result for the effective free energy:
\begin{equation}
-k_BT  \ln\left(\sum_k |z_k|^{-N} \right)
\label{eq:free1}
\end{equation}
If $N$ is large enough that the $z_k$ closest to the origin dominates, then the free energy reduces to
\begin{equation}
Nk_BT \ln z_1
\label{eq:free2}
\end{equation}
where $z_1$ is the location of the singularity that lies closest to the origin.

\subsection{Pressure in the absence of a potential energy: scaling formulas}

 In the case of an interval with no potential, we can write
\begin{equation}
z_k=-z_c-w\left(\frac{k\pi}{L} \right)^2
\label{eq:zvals}
\end{equation}
where $z_c=w=1/4$. Then, the free energy has the form
\begin{eqnarray}
\lefteqn{-k_BT \ln \left[ \sum_k  \left(z_c+w\left(\frac{k\pi}{L} \right)^2\right)^{-N}\right] } \nonumber \\ &=& Nk_BT \ln z_c - k_BT \ln \left[ \sum_k  \left(1+\frac{w}{z_c}\left(\frac{k\pi}{L} \right)^2\right)^{-N}\right] \nonumber \\ & \rightarrow & Nk_BT \ln z_c - k_BT \ln \left[ \sum_ke^{-N\frac{w}{z_c}(k \pi/L)^2} \right]
\label{eq:free3}
\end{eqnarray}
The last line of (\ref{eq:free3}) follows if $L$ is sufficiently large. We assume this to be the case and proceed. The pressure is the negative of the derivative of the free energy with respect to $L$. Making use of (\ref{eq:free3}) we find
\begin{equation}
P = 2 k_BT \frac{Nw}{z_c} \frac{\sum_k \frac{k^2 \pi^2}{L^3} e^{-N\frac{w}{z_c}(k \pi/L)^2}}{\sum_ke^{-N\frac{w}{z_c}(k \pi/L)^2}}
\label{eq:press1}
\end{equation}

\begin{figure}[htbp]
\begin{center}
\includegraphics[width=3in]{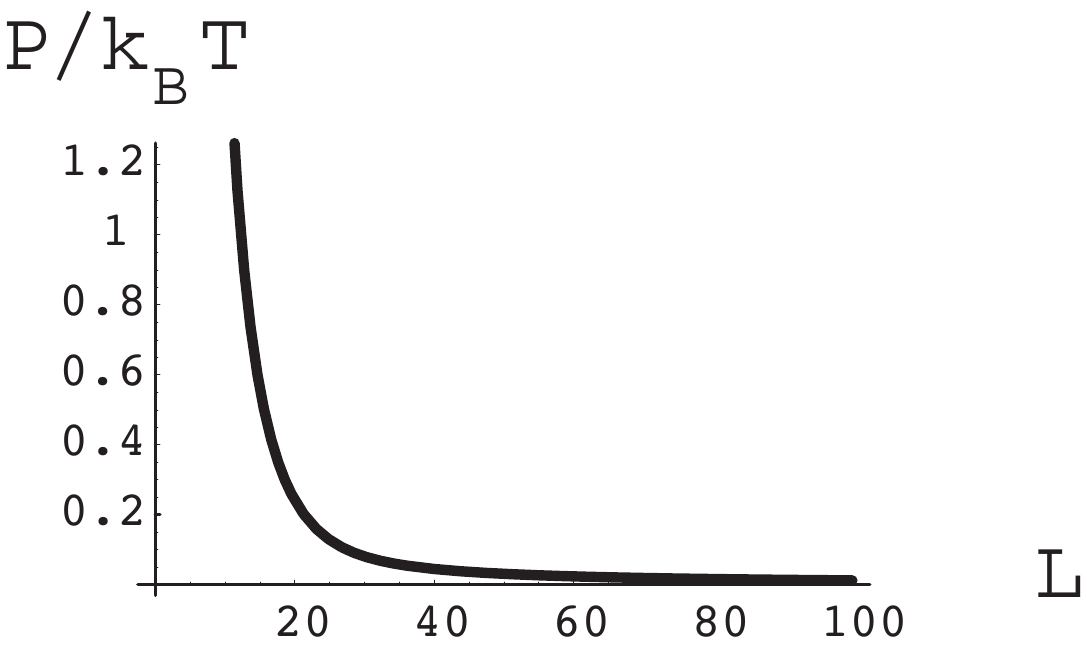}
\caption{The pressure as a function of $L$ when $N=100$. Here, $w=z_c=1$.  }
\label{fig:pressureplot1}
\end{center}
\end{figure}

Keeping the full expression for the free energy, we obtain a general result for the force between the two walls. Figure \ref{fig:pressureplot1} illustrates the dependence on $L$ of the force divided by $k_BT$, with $w=z_c=1$ and $N=100$.
One can easily show that at sufficiently large $L$ the pressure will go as $1/L$.

\begin{figure}[htbp]
\begin{center}
\includegraphics[width=3in]{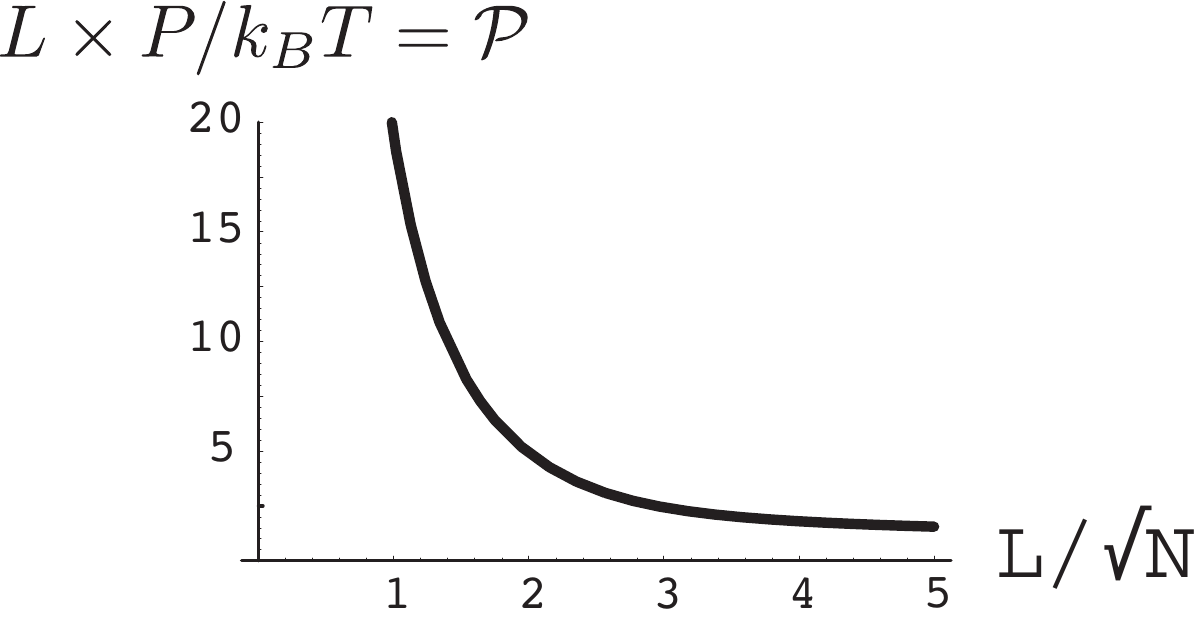}
\caption{The quantity $L P(L,N)/k_BT = \mathcal{P}(L/\sqrt{N})$.}
\label{fig:pressureplot2}
\end{center}
\end{figure}

In fact, a straightforward analysis tells us that the pressure will, in this case have the general form
\begin{equation}
P(L,N)/k_BT =  \mathcal{P}(L/\sqrt{N})/L
\label{eq:genpress1}
\end{equation}
Figure \ref{fig:pressureplot2} is a graph of the function on the right hand side of (\ref{eq:genpress1}).
The large $L$ behavior of the quantity $\mathcal{P}$ is evident from the figure, in that it approaches a limiting value as its argument goes to infinity.

\begin{figure}[htbp]
\begin{center}
\includegraphics[width=3in]{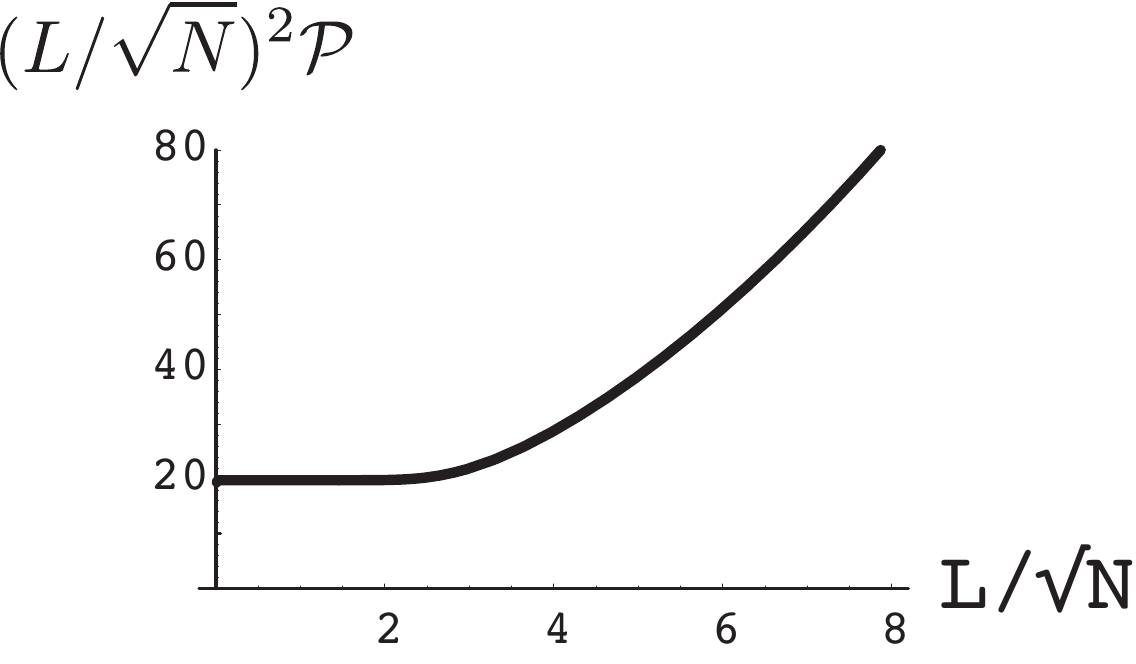}
\caption{ The quantity $(L.\sqrt{N})^2 \mathcal{P}$, plotted against $L/\sqrt{N}$. }
\label{fig:pressureplot3}
\end{center}
\end{figure}

There is also the question of the behavior of the pressure at intermediate values of $L\sqrt{N}$. In the regime in which one singularity in $z$ dominates, one can show that the pressure goes as $N/L^3$. This is consistent with $\mathcal{P} \propto N/L^2$. Figure \ref{fig:pressureplot3} is a plot of $(L/\sqrt{N})^2 \times \mathcal{P}$.

Note that this combination is effectively a constant between $L/\sqrt{N}=0$ and $L/\sqrt{N}=3$.  From this we can infer two different regimes for the pressure. In the first, when $L \gg \sqrt{N}$, the force goes as $1/L$, independent of $N$. In the second in which $L/\sqrt{N} \simeq 1$, the force goes as $N/L^3$.

\section{The influence of attractive potentials}

\begin{figure}[htbp]
\begin{center}
\includegraphics[width=2in]{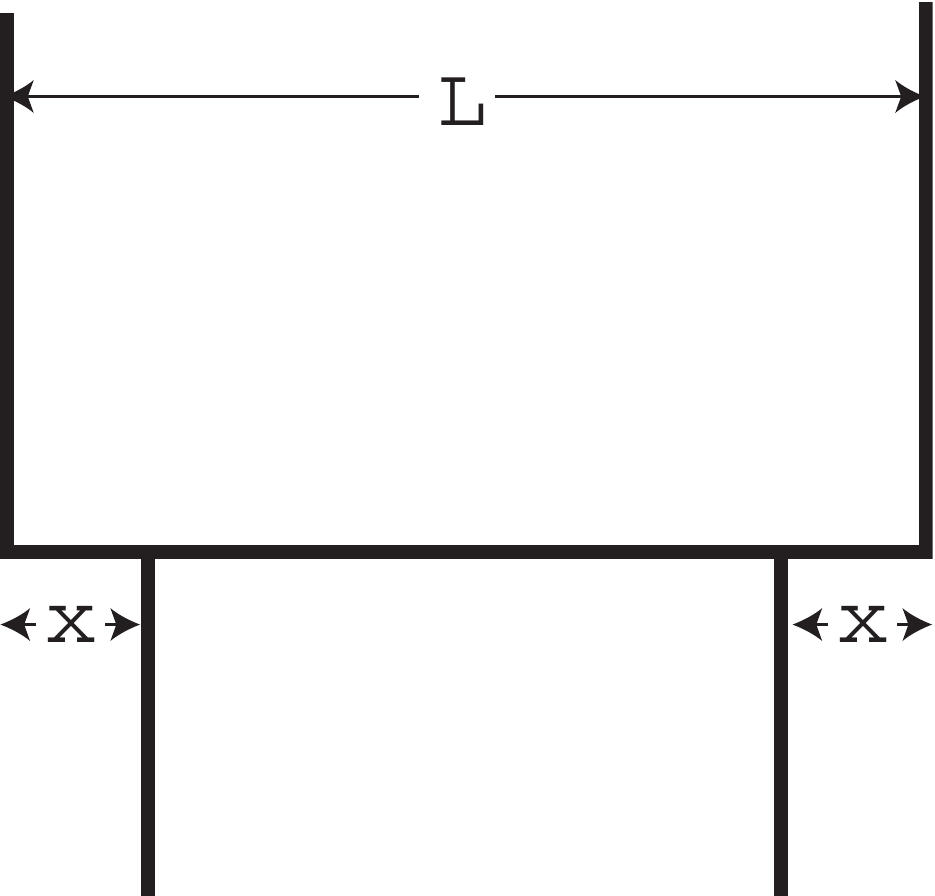}
\caption{The potential configuration with which we will work. There are two attractive delta function potentials, each a distance $x$ from the bounding walls of a region with an extent $L$. }
\label{fig:pots}
\end{center}
\end{figure}

Once again, we assume that there are attractive potentials near the two bounding surfaces, as indicated in Fig. \ref{fig:pots}.
There will be two types of eigenfunction in this configuration: even and odd parity about the center. In the region $0<y<L/2$, the even functions have the form
\begin{equation}
\psi_e(y) = \left\{ \begin{array}{ll} \sin k y & y<x \\ \cos(k(L/2-y)) & x<y<L/2 \end{array} \right.
\label{eq:psie}
\end{equation}
The odd parity functions are of the form
\begin{equation}
\psi_o(y) = \left\{ \begin{array}{ll} \sin k y & y<x \\ \sin(k(L/2-y)) & x<y<L/2 \end{array} \right.
\label{eq:psio}
\end{equation}
The equation for the eigenvalues, expressed through the quantity $k$, is
\begin{equation}
\frac{\psi^{\prime}(x^-)}{\psi(x^-)} - \frac{\psi^{\prime}(x^+)}{\psi(x^+)} = V
\label{eq:kcond1}
\end{equation}
where $V$ is the strength of the delta function potential,  $x^-$ is just to the left of $x$ and $x^+$ is just to the right. Making use of the two forms in (\ref{eq:psie}) and (\ref{eq:psio}), we have for the equations satisfied by $k$ in the case of the even and odd eigenfunctions, respectively
\begin{eqnarray}
k \cot (k x)+k \tan \left(k
   \left(x-\frac{L}{2}\right)\right)&=&V \label{eq:eveneq} \\
k \cot (k x)-k \cot \left(k
   \left(x-\frac{L}{2}\right)\right)&=&V \label{eq:oddeq}
\end{eqnarray}
For sufficiently large values of $V$, there may also be solutions at imaginary $k = i \kappa$. The equations that are satisfied in the even and odd parity cases are, respectively,
\begin{eqnarray}
\kappa \coth (\kappa x)-\kappa \tanh \left(\kappa
   \left(x-\frac{L}{2}\right)\right) &=&V \label{eq:eveneq1} \\
   \kappa \coth (\kappa x)-\kappa \coth \left(\kappa
   \left(x-\frac{L}{2}\right)\right) &=&V \label{eq:oddeq1}
\end{eqnarray}
The threshold values of $V$ for which there are solutions to (\ref{eq:eveneq1}) and (\ref{eq:oddeq1}) are, respectively
\begin{eqnarray}
V_{\rm even} & = & \frac{1}{x} \label{eq:veven} \\
V_{\rm odd} & = & \frac{1}{x} + \frac{1}{L/2-x} \label{eq:vodd}
\end{eqnarray}

Given these solutions, we are able construct the expression for the force generated by the rooted branched polymers.  As previously, we start with the result (\ref{eq:free1}) for the effective free energy. The general result for the force associated with this free energy, corresponding to (\ref{eq:press1}), is
\begin{eqnarray}
\frac{P}{k_BT} & = & \frac{\partial}{\partial L} \ln \sum_l |z_l|^{-N} \nonumber \\
& = & \frac{-N\sum_l |z_l|^{-N-1}\partial |z_l| / \partial L}{\sum_l  |z_l|^{-N}}
\label{eq:press2}
\end{eqnarray}
To assess the derivative of the singularities with respect to $L$, we make use of the general relationship
\begin{equation}
|z_l| =  \frac{1}{4 \cos ^2 k_l}
\label{eq:zonk}
\end{equation}
Then,
\begin{equation}
\frac{\partial |z_l|}{\partial L} = -\frac{\sin k_l}{2 \cos^3 k_l} \frac{\partial k_l}{\partial L}
\label{eq:zonL}
\end{equation}
There will be corresponding equations for the up to two bound states, in which an analytic continuation has been performed from the variable $k$ to the variable $i \kappa$.

\begin{figure}[htbp]
\begin{center}
\includegraphics[width=3in]{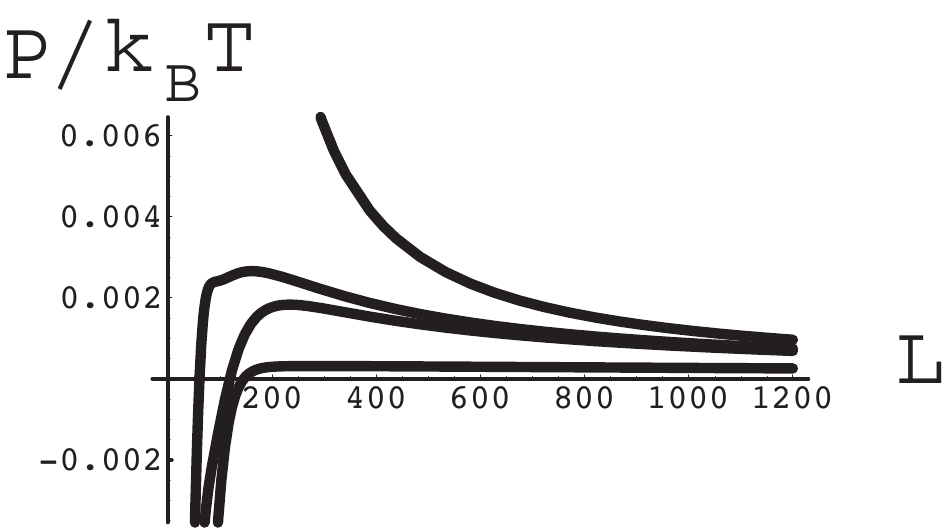}
\caption{The pressure as a function of distance, $L$, between the walls for $x=0.1$ and the following values of the attractive potential strength, $V$: 9.9, 10.01, 10.02, 10.05. The smaller $V$ the higher the pressure. }
\label{fig:vpressplot1}
\end{center}
\end{figure}

Equations (\ref{eq:press2})--(\ref{eq:zonL}) allow us to calculate results for the force exerted by the branched polymer between two walls. Figure \ref{fig:vpressplot1} summarizes results for the pressure for various values of the attractive potential. In the case of interest here, there is no bound state unless $V>10$.
Note that for sufficiently small separations and sufficiently attractive potentials the pressure becomes negative, corresponding to an attraction between the walls. This follows from the fact that the bound state free energy decreases as the attractive wells approach each other, in analogy to the simplest version of the chemical bond \cite{pauling}.

\begin{figure}[htbp]
\begin{center}
\includegraphics[width=3in]{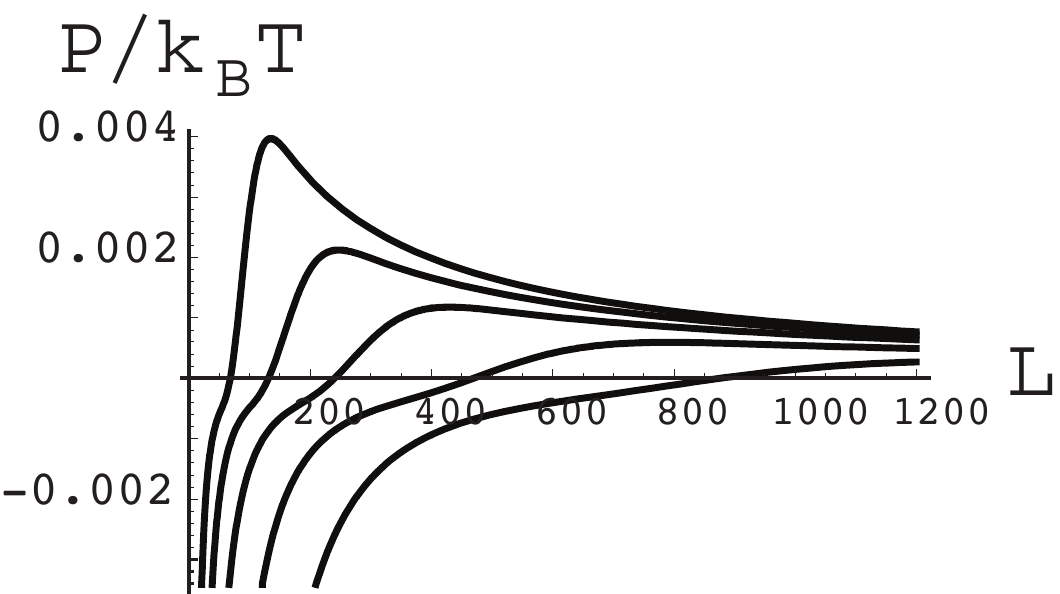}
\caption{The pressure divided by $k_BT$ for $V=10.00025$ and the following values of $N$: 300,000, 100,000, 30,000, 10,000, 3,000. The larger $N$ the more negative (or less positive) the pressure. }
\label{fig:pressurebound}
\end{center}
\end{figure}

\begin{figure}[htbp]
\begin{center}
\includegraphics[width=2.5in]{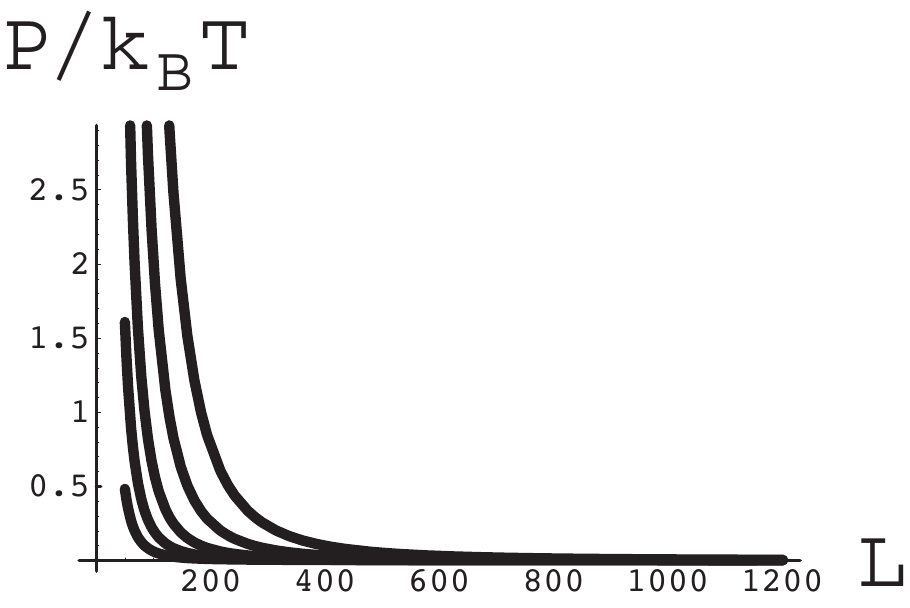}
\caption{The pressure divided by $k_BT$ for $V=9.9997$ and the following values of $N$: 300,000, 100,000, 30,000, 10,000, 3,000. Here, the larger $N$ the greater (more positive) the pressure. }
\label{fig:pressureunbound}
\end{center}
\end{figure}

Figures \ref{fig:pressurebound} and \ref{fig:pressureunbound} are two interesting and contrasting plots. In both figures, the attractive potential is a distance 0.1 from the edges of the system. The threshold for a bound state is $V=10$. In the first plot, $V=10.00025$, which is just sufficiently strong that there is a bound state In the second plot, $V=9.9997$, and the attractive delta function does not quite suffice to produce such a solution to the Schr\"{o}dinger-like equation. The plots are for a range of values of $N$, as indicated in the caption. In the case of the first plot, Fig. \ref{fig:pressurebound}, the larger the value of $N$, the lower, or more negative, the pressure. In the case of the second plot, Fig. \ref{fig:pressureunbound}, the greater $N$, the higher the pressure.

What this tells us is a very small change in the attractive potential suffices to give rise to a considerable change in the force between the walls. It also tells us that an attractive potential can be ``mediated'' by the branched polymer in such a way as to facilitate the assembly of a capsid.

Finally, note in the case of Fig. \ref{fig:pressurebound} that the force is concave downward when $L$ is small enough. This leads to a mechanical {\em instability} in that range, in which the walls will continue to collapse, assuming a constant countervailing force. This means that the walls will be pulled together until other mechanisms, such as excluded volume effects, intervene.


\begin{acknowledgments}
The authors of the paper gratefully acknowledge useful conversations with David Schwab. J.R. is especially indebted to Christian Rose for collaborations on a project that led directly to some of the key elements reported in this paper. This research was supported by the National Science Foundation through DMR Grant 04-04507.
\end{acknowledgments}

\appendix

\section{Normalization of eigenfunctions} \label{app:norm1}

We start with the two equations
\begin{eqnarray}
-\frac{d^2\phi_1(x)}{dx^2} + V(x) \phi_1(x) & = & E \phi_1(x) \label{eq:pp1} \\
-\frac{d^2 \phi_2(x)}{dx^2} + V(x) \phi_2(x) & = & (E + \Delta E) \phi_2(x) \label{eq:pp2}
\end{eqnarray}
We will assume that $\phi_1(x)$ has zeros at $x=0$ and $x=x_0$. As for $\phi_2(x)$, it is also zero at $x=0$, and it has another zero close to $x_0$. In fact, we assume that as $\Delta E \rightarrow 0$, $\phi_2(x) \rightarrow \phi_1(x)$. Subtracting (\ref{eq:pp1}) from (\ref{eq:pp2}) and integrating from $x=0$ to $x=x_0$, we find
\begin{eqnarray}
\lefteqn{\int_{0}^{x_0} \left[ \phi_2(x) \frac{d^2 \phi_1(x)}{dx^2} - \phi_1(x) \frac{d^2 \phi_2(x)}{dx^2}\right] dx} \nonumber \\
& = & \int_0^{x_0} \frac{d}{dx} \left[ \phi_2(x) \frac{d\phi_1(x)}{dx} - \phi_1(x) \frac{d\phi_2(x)}{dx}\right]dx \nonumber \\
& = & \left. \left[ \phi_2(x) \frac{d\phi_1(x)}{dx} - \phi_1(x) \frac{d\phi_2(x)}{dx}\right] \right|_0^{x_0} \nonumber \\
& = & \Delta E \int_0^{x_0} \phi_1(x) \phi_2(x) dx
\label{eq:pp3}
\end{eqnarray}
As for the next to last line in (\ref{eq:pp3}), because of boundary conditions all contributions are equal to zero except $\phi_2(x_0) (d \phi_1(x)/dx) |_{x=x_0}$. Under the assumption that $\Delta E$ is small, the last line reduces to $\Delta E \int_0^{x_0} \phi_1(x)^2 dx$. That is, we now have
\begin{equation}
\phi_2(x_0)\left. \frac{d\phi_1(x)}{dx}\right|_{x=x_0} = \Delta E \int_0^{x_0} \phi_1(x)^2 dx
\label{eq:pp4}
\end{equation}
Given that the zero of $\phi_2(x)$ is close to $x_0$, we can write
\begin{equation}
\phi_2(x_0) = \left( \left.- \frac{d \phi_1(x)}{dx}\right|_{x=x0} \right) \frac{dx_0}{dE} \Delta E
\label{eq:pp5}
\end{equation}
This can be established graphically. Combining the results, we have
\begin{equation}
- \left( \left. \frac{d\phi_1(x)}{dx} \right|_{x=x_0} \right)^2 \frac{dx_0}{dE} = \int_0^{x_0} \phi_1(x)^2 dx
\label{eq:pp6}
\end{equation}

Now, let us look at the denominator $\phi_n \phi_{n-1}$ in (\ref{eq:gencase9}). In the vicinity of the zero of $\phi_n$, we can write
\begin{eqnarray}
\phi_n &= &\phi(n-x_0(z))
\label{eq:pp7}
\end{eqnarray}
where
\begin{equation}
x_0(z) = n + \frac{dx_0}{dz} \Delta z
\label{eq:pp8}
\end{equation}
On the other hand
\begin{eqnarray}
\phi_{n-1} & =  & \phi(-1 - \Delta z \  dx_0/dz ) \nonumber \\
& \rightarrow& - \frac{d \phi_n}{dn}
\label{eq:pp9}
\end{eqnarray}
This means that we have
\begin{eqnarray}
\phi_n \phi_{n-1} & = & \phi(- \Delta z \ dx_0/dz) \phi(-1) \nonumber \\
& \rightarrow & \left( \frac{d \phi_n}{dn}\right)^2 \frac{dx_0}{dz} \Delta z
\label{eq:pp10}
\end{eqnarray}
If we divide by this, we have a contribution to the residue of the pole going exactly like $1/(\int_0^n \phi_n^2 \ dn)$, to within a multiplicative constant. In fact, a careful analysis of the relationship between the variables $z$ in our system and the energy, $E$, in the Schr\'{o}dinger equation leads to the conclusion that the multiplicative constant is precisely 1/4. This means that, to within the multiplicative factor of 1/4, the residue effectively normalizes the contributions to the density.

\section{Normalization in the discrete system} \label{app:normalizationdis}
We start with the following version of the equation for the quantity $\phi_n$
\begin{equation}
e^{- \beta u_{n+1}/2} \phi_{n+1}(k) + e^{- \beta u_n/2} \phi_{n-1}(k) = 2 \cos k  \  \phi_n(k)
\label{eq:neq1}
\end{equation}
We consider two versions of this equation, one just like (\ref{eq:neq1}) and one with the parameter $k$ slightly different. Denoting that new value of $k$ as $k^{\prime}$, we have, multiplying (\ref{eq:neq1}) by $\phi_n(k^{\prime})$ and the corresponding equation for $k=k^{\prime}$ by $\phi_n(k)$, subtracting the results and summing over $n$,
\begin{widetext}
\begin{eqnarray}
\lefteqn{\sum_{n=n_b}^{n_e} \phi_n(k^{\prime}) \left( e^{- \beta u_{n+1}/2} \phi_{n+1}(k) + e^{- \beta u_n/2} \phi_{n-1}(k) \right) } \nonumber \\
&&-\sum_{n=n_b}^{n_e} \phi_n(k) \left( e^{- \beta u_{n+1}/2} \phi_{n+1}(k^{\prime}) + e^{- \beta u_n/2} \phi_{n-1}(k^{\prime}) \right) \nonumber \\
& = & \sum_{n=n_b}^{n_e} \left( \phi_n(k^{\prime}) e^{- \beta u_{n+1}/2} \phi_{n+1}(k) - \phi_n(k) e^{- \beta u_n/2} \phi_{n-1}(k^{\prime})\right) \nonumber \\
&& -  \sum_{n=n_b}^{n_e} \left( \phi_n(k) e^{- \beta u_{n+1}/2} \phi_{n+1}(k^{\prime}) - \phi_n(k^{\prime}) e^{- \beta u_n/2} \phi_{n-1}(k)\right) \nonumber \\
& = & \phi_{n_e}(k^{\prime}) e^{- \beta u_{n_e+1}/2} \phi_{n_e+1}(k) - \phi_{n_b}(k) e^{- \beta u_{n_b}/2} \phi_{n_b-1}(k^{\prime}) \nonumber \\
&& -\phi_{n_e}(k) e^{- \beta u_{n_e+1}/2} \phi_{n_e+1}(k^{\prime}) + \phi_{n_b}(k^{\prime}) e^{- \beta u_{n_b}/2} \phi_{n_b-1}(k) \nonumber \\
& = & 2( \cos k - \cos k^{\prime}) \sum_{n=n_b}^{n_e} \phi_n(k^{\prime}) \phi_n(k)
\label{eq:neq2}
\end{eqnarray}
\end{widetext}
We will take the solutions of the equations in the equation above to satisfy the boundary conditions in Section \ref{sec:newmethod}. Furthermore, we take the end point of the summation to be $n_b=1$, $n_e=L-1$. Furthermore, we will assume that the potential energy is zero at and near the boundaries. Then, $\phi_{n_b}(k) = \phi_{n_b-1}(k)$ and similarly for $k^{\prime}$. Additionally, we will assume that the value $k$ is consistent with the boundary condition $\phi_{n_e=L}(k)=0$. Then, (\ref{eq:neq2}) reduces to
\begin{equation}
- \phi_{L-1}(k) \phi_{L}(k^{\prime}) = 2(\cos k - \cos k^{\prime}) \sum_{n=1}^{L-1} \phi_n(k) \phi_n(k^{\prime})
\label{eq:neq3}
\end{equation}

From here on, the analysis follows that in Appendix \ref{app:norm1}.

\section{Note on the normalization of extended eigenstates} \label{app:normalization}

As an essential step in the calculation of generating functions, we establish the proper way to normalize the eigenfunctions we deal with in the case of very large systems. In particular, we are interested in the case of an eigenfunction in a long interval that goes as $\phi(x) \propto \sin(kx+ \theta(k))$ towards the left end of the interval. The boundary condition will be $\phi(x_0)=0$, which tells us that
\begin{equation}
k_lx_0 + \theta(k_l) = l \pi
\label{eq:norm1}
\end{equation}
Recall (\ref{eq:pp6}). For a given value of $k_l$, we have
\begin{equation}
\Delta x( x_0 + \theta^{\prime}(k_l)) + x_0\Delta k =0
\label{eq:norm2}
\end{equation}
This tells us that
\begin{equation}
\frac{dx_0}{dk} = - \frac{x_0 + \theta^{\prime}(k)}{k}
\label{eq:norm3}
\end{equation}
Given that, in our version of the Schr\"{o}dinger equation, $E=k^2$, we can rewrite (\ref{eq:norm3}) as
\begin{eqnarray}
\frac{dx_0}{dE} & = & \frac{1}{2k}\frac{dx_0}{dk} \nonumber \\
& = & - \frac{x_0 + \theta^{\prime}(k)}{2k^2}
\label{eq:norm4}
\end{eqnarray}
From (\ref{eq:pp6}) and the above, we have
\begin{equation}
\int_0^{x_0} \phi(x)^2 dx = \left(\frac{d \phi(x)}{dx} \right)_{x=x_0}^2 \frac{x_0 + \theta^{\prime}(k)}{2k^2}
\label{eq:norm5}
\end{equation}

We now return to (\ref{eq:norm1}). If the integer $l$ increments by one, then $k$ will change as follows
\begin{equation}
\Delta k(x_0 + \theta^{\prime}(k)) = \pi
\label{eq:norm6}
\end{equation}
This tells us that
\begin{equation}
\Delta k \frac{(x_0 + \theta^{\prime}(k))}{\pi} = 1
\label{eq:norm7}
\end{equation}
and, we then have the following result for the eigenfunction sum leading to the density:
\begin{eqnarray}
\lefteqn{\sum_l \frac{\phi_l(x)^2}{\int _0^{x_0} \phi_l(x)^2 dx}} \nonumber \\ & = & \sum_l \phi_l(x)^2 \frac{2k_l^2}{\left( d \phi(x)/dx\right)^2_{x=x_0} (x_0 + \theta^{\prime}(k_l))} \frac{\Delta k (x_0 + \theta^{\prime}(k_l))}{ \pi} \nonumber \\
& \rightarrow& \frac{2}{\pi} \int \frac{k^2}{\left( d \phi(x)/dx\right)^2_{x=x_0}} \phi(x)^2 dk
\label{eq:norm8}
\end{eqnarray}
If we choose $\phi(x)$ \emph{precisely} equal to $\sin(kx+ \theta(k))$ at $x$ near $x_0$, then the last line in (\ref{eq:norm8}) reduces to
\begin{equation}
\frac{2}{\pi} \int \phi(x)^2 dk
\label{eq:norm9}
\end{equation}

\section{Reconstruction of the sum $\sum_l \phi_l(x)^2$.} \label{app:sumcalc}

We focus on the case $x<x_1$ and begin with the extended states, and we will seek the difference between this sum and the sum in the absence of the attractive potential. That is, our task is to find the value of the sum.
\begin{equation}
\sum_k \frac{\sin^2 kx}{ \sin^2kx_1} \left[ \sin^2 (kx_1 + \theta(k)) - \sin^2(kx_1) \right]
\label{eq:sd8}
\end{equation}
After expanding $\sin(kx_1 + \theta)$ and carrying out some simple algebra and trigonometry, the above expression reduces to
\begin{equation}
\sum_k \frac{\sin^2 kx}{\sin^2 kx_1} \times \left(-\frac{1}{2}\right)  \mathop{\rm Re} \left[ e^{2ikx_1}\left(e^{2i \theta(k)} -1\right)\right]
\label{eq:sd9}
\end{equation}
Given the trigonometric identity
\begin{equation}
e^{2i \theta} = \frac{i \tan \theta + 1}{1-i \tan \theta}
\label{eq:sd10}
\end{equation}
The summand becomes
\begin{eqnarray}
\lefteqn{-\frac{1}{2} \frac{\sin^2 kx}{\sin^2 kx_1} \mathop{\rm Re} \left[ e^{2ikx_1} \frac{2i \tan \theta(k)}{1-i \tan \theta(k)} \right]} \nonumber \\ & = &  -\sin^2 kx \mathop{\rm Re} \left[ \frac{i e^{2ikx_1} V}{k-V \sin k x_1e^{ikx_1}}\right]
\label{eq:sd11}
\end{eqnarray}
where (\ref{eq:sd7}) has been used. The integral to be performed is
\begin{equation}
-\frac{1}{\pi}\int_{-\infty}^{\infty}\sin^2 kx \mathop{\rm Re} \left[ \frac{i e^{2ikx_1} V}{k-V \sin k x_1e^{ikx_1}}\right] dk
\label{eq:sd12}
\end{equation}
where the removal of a factor of two in front of the integral (see  (\ref{eq:norm9})) is compensated for by the fact that the range of integration has been extended by a factor of two. In fact, the result of the integration will be real as a matter of course, so we can remove the ``Re'' function from the expression. The sole contribution to the integration is  from a pole on the upper imaginary axis, at a value of $k=i\kappa$ such that the denominator $k-V \sin kx_1 e^{ikx_1} \rightarrow i( \kappa - V \sinh \kappa x_1e^{-\kappa x_1}) =0$. This equation for the pole is the same as the requirement (\ref{eq:sd4}) for the bound state. The residue at that pole is\begin{equation}
\frac{1}{\pi} \times 2 \pi i \times \frac{iV \sinh^2 \kappa x e^{-2 \kappa x_1}}{\left(\frac{d}{dk}(k-V \sin kx_1e^{ikx_1})\right)_{k=i \kappa}}
\label{eq:sd13}
\end{equation}
Focusing on the denominator in (\ref{eq:sd13}), we have
\begin{eqnarray}
\lefteqn{\left(\frac{d}{dk}(k-V \sin kx_1e^{ikx_1})\right)_{k=i \kappa} } \nonumber \\ &=& \left(1- Vx_1e^{2ikx_1}\right)_{k=i \kappa} \nonumber \\
& = & 1-Vx_1e^{-2 \kappa x_1} \nonumber \\
& = & 1-\frac{\kappa x_1 e^{- \kappa x_1}}{\sinh \kappa x_1} \nonumber \\
& = & \frac{e^{- \kappa x_1}}{\sinh \kappa x_1} \left( \sinh \kappa x_1 e^{\kappa x_1}  - \kappa x_1\right)
\label{eq:sd14}
\end{eqnarray}
The third line of (\ref{eq:sd14}) follows from (\ref{eq:sd4}). Inserting this result into (\ref{eq:sd13}) we end up with the result for the integration
\begin{eqnarray}
\lefteqn{-\frac{2V e^{- \kappa x_1} \sinh \kappa x_1}{\sinh \kappa x_1 e^{\kappa x_1} - \kappa x_1} \sinh^2 \kappa x} \nonumber \\ &=& - \frac{2 \kappa }{\sinh \kappa x_1 e^{\kappa x_1} - \kappa x_1} \sinh^2 \kappa x
\label{eq:sd15}
\end{eqnarray}
Again, we have utilized (\ref{eq:sd4}) to obtain the right hand side of the equation above. Referring to Section \ref{subsec:boundstate}, we see that this exactly cancels the contribution of the bound state to the left of the delta function potential. We have recovered the standard result for a complete orthonormal set of eigenfunctions, and that is that the sum is independent of the set chosen.

The corresponding integration when $x>x_1$ is
\begin{equation}
-\frac{1}{\pi} \int_{-\infty}^{\infty} \sin^2 k x_1 \mathop{\rm Re} \left[ \frac{ie^{2ikx}V}{k-V \sin kx_1e^{ikx_1}}\right] dk
\label{eq:sd16}
\end{equation}
The same sort of contour integration yields a result that precisely cancels the contribution of the bound state in that regime.

\section{dimensional reduction for a spherical geometry}\label{app:curved}

A prime motivation for the work reported here is the packing of complex RNA into a viral capsid. The closest approximation to the geometry of the packing environment entails spherical symmetry. It is natural to ask whether dimensional reduction will prove useful in this case. In this Appendix, we explore the consequences of the results of Brydges and Imbrie \cite{bi1,bi2} when spherical symmetry holds.

\begin{figure}[htbp]
\begin{center}
\includegraphics[width=3in]{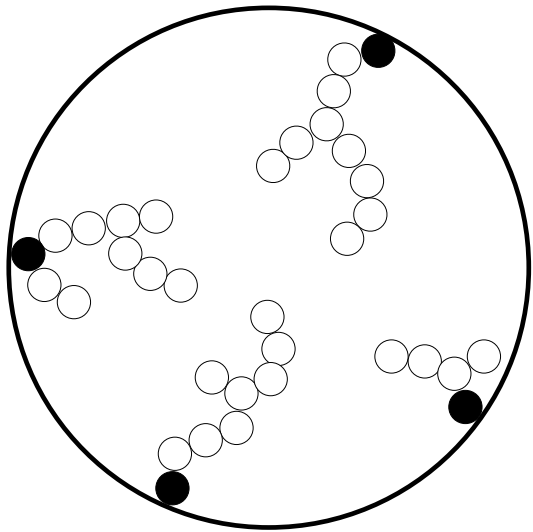}
\caption{Solution of branched polymers. Roots are depicted by solid black circles.}\label{fig:BPsolution}
\end{center}
\end{figure}

The two physical systems related by dimensional reduction are a hard-core classical gas in $D$ dimensions and a solution of rooted branched polymers in $d=D+2$ dimensions. The particles have a hard core repulsive interaction with diameter $a$, and they are confined in a container with radius $R$. The equation relating these two systems is:
\begin{equation}
G_{HG}(z)=G_{BP}(-\frac{z}{2\pi a}, \frac{z}{2\pi R})
\label{eq:main-spherical}
\end{equation}
where $G_{HG}$ on the left side of the final equation is the grand partition function of the gas, $z$ being the fugacity of the particles. The quantity $G_{BP}(z,w)$, on the right side of the equation, is the grand partition function of the solution of annealed branched polymers. The first argument, $z$, is the fugacity of the monomers and the second argument, $w$, is the fugacity of the roots. The roots are constrained to be on the surface of the container (see figure \eqref{fig:BPsolution}).

We provide the proof of the equation \eqref{eq:main-spherical} in the following sections. The reasoning is parallel to the one suggested by Cardy \cite{cardy1}. We start from the partition function of hardcore classical gas in. Then we continue with defining the partition function of a super gas. We show those two are equal using cluster expansion and the properties of gaussian integration. Finally we put the partition function of a super gas into the form of the partition function of a solution of branched polymers. This last step is done using the Taylor expansion of functions of Grassman variables.

\subsection{Classical Hardcore Gas}
Consider the classical gas in a $D$ dimensional space. We assume the interaction potential energy of two gas particles, $U(r_{ij}^2)$, depends only on the distance between those two particles: $r_{ij}^2$. The labels, $i$ and $j$, indicate the gas particles. We also include an external potential, $V(r_i ^2)$. This potential depends on the radial distance of the gas particle from the origin: $r_i ^2$. The partition functions of the hardcore gas is
\begin{eqnarray}
G_{HG}(z)&=&\sum{Z_{HG}(N)} z^N \\
Z_{HG}(N)&=&\frac{1}{N!} \int \prod_i \ud^D r_i e^{-\sum_{ij}{U(r_{ij}^2)}-\sum_i{V(r_i^2)}} \nonumber \\
\label{eq:pfunction-chg}
\end{eqnarray}
If we define:
$$g_{ij}=g_{r_{ij}^2}=\exp(U(r_{ij}^2))-1$$
and

$$h_i=h_{r_i^2}=\exp(V(r_i^2))$$
then it is well known  that the grand partition function can also be represented in the form of cluster expansion \cite{Huang}.
\begin{eqnarray}
G_{HG}(z)&=&\exp(\sum_k{b_k z^k})\\
b_k&=&\frac{1}{k!} \sum_{\mathcal{CG}}{ \int{\prod_i^k{\ud^D r_i h_i} \prod_{ij}g_{ij}}}\\
\label{eq:cluster-expansion}
\end{eqnarray}
The label $\mathcal{CG}$ stands for 'Connected Graphs'. $b_k$ is the sum of all connected graphs (clusters) with $k$ particles divided by $k!$. in the next section we define the partition function of a super symmetric gas.

\subsection{supersymmetric gas}
Consider the supersymmetric gas in a super-space of $d=D+2$ real plus 2 Grassman coordinates. As in Cardy's exposition \cite{cardy1} the distance between two particles and also a particle from the origin is defined as:
\begin{eqnarray}
R_{ij}^2 &=& (r_i-r_j)^2+(\thetab_i-\thetab_j)(\theta_i-\theta_j)\\
R_i^2 &=& r_i^2+\thetab_i \theta_i
\label{eq:ss-distance}
\end{eqnarray}
The quantities $r_{ij}$ and $r_i$ are the usual real-valued distances in $d$ dimensional real space, while $\theta_i$ and $\thetab_i$ are the Grassman coordinates of the particles. As in the case of the classical hardcore gas, it is natural to define the partition function of a supersymmetric gas as (`SSG' stands for `Super Symmetric Gas'):
\begin{widetext}
\begin{eqnarray}
G_{SSG}(z)&=&\sum{Z_{HG}(N)} z^N \\
Z_{SSG}(N)&=&\frac{1}{N!} \int \prod_i \ud^d r_i \ud \thetab_i \ud \theta_i e^{-\sum_{ij}{U(R_{ij}^2)}-\sum_i{V(R_i^2)}}
\label{eq:pfunction-ssg}
\end{eqnarray}
\end{widetext}
Here, the cluster expansion works as well. The only modification is that wherever we have $r_{ij}^2$ or $r_i^2$ in the classical case, we use their counterparts for the supersymmetric case, i.e. $R_{ij}^2$ and $R_i^2$. In the next section we demonstrate that, given a supersymmetric gas in super-space of $d=D+2$ real and two Grassman coordinates and a hardcore classical gas in a $D$ dimensional space with the same inter-particle and external potentials, we have
\begin{equation}
G_{SSG}(z)=G_{HG}(z)
\label{eq:ssg-hg}
\end{equation}

\subsection{Super Symmetric Gas versus Classical Gas}
As in \cite{cardy1}, we define the functions $p(\mu)$ and $q(\nu)$ as follows:
\begin{eqnarray*}
g(r_{ij}^2)&=&\int_0^\infty \ud \mu_{ij} p(\mu_{ij}) e^{-\mu_{ij}r_{ij}^2}\\
h(r_i^2)&=&\int_0^\infty \ud \nu_{i} q(\nu_{i})e^{-\nu_{i}r_{i}^2}
\end{eqnarray*}
Notice that in the above expressions, $R_{ij}^2$ can also be used instead of $r_{ij}^2$. Consider the contribution of of a connected graph of $k$ supersymmetric particles, $\mathcal{CG}$. We use the above relations to transform the functions $g$ and $h$ into the functions $p$ and $q$. Regardless of integrations over the parameters $\mu_{ij}$ and $\nu_i$, we are left with
$$\int \prod_{i=1}^k \ud^d r_i \ud \thetab_i \theta_i \exp(-\frac{1}{2} \sum_{jl}\big(r_j A_{jl}^{\mathcal{CG}} r_l + \thetab_j A_{ij}^{\mathcal{CG}} \theta_l \big)$$
The quantity $A^{\mathcal{CG}}$ is a $k\times k$ matrix dependent on the connected graph $\mathcal{CG}$. The gaussian integration over Grassman numbers evaluates to
$$(-\frac{1}{2} \Lambda)^k \det(A^{\mathcal{CG}})$$
We have used the following convention in this evaluation:
$$\int \ud \thetab \ud \theta \ \thetab \theta=\Lambda $$
On the other hand the integration over the real coordinates contribute a factor of
$$(2 \pi)^{kd/2} \det(A^{\mathcal{CG}})^{-d/2}$$
If we choose $\Lambda=-\frac{1}{\pi}$, the final expression evaluates to
$$(2 \pi)^{k(d-2)/2} \det(A^{\mathcal{CG}})^{(d-2)/2}$$
The two Grassman coordinates have canceled the effect of two real coordinates. It is evident that if we start with a classical hardcore gas in $D=d-2$ dimensions, we arrive at the same expression as the above.

This completes the proof for the equation \eqref{eq:ssg-hg}.

In the next section, we expand the partition function of a supersymmetric gas. However, instead of cluster expansion, we use a Taylor expansion. It turns out that the super-symmetric gas is related to the generating function of branched polymers.

\subsection{Super Symmetric Gas versus Branched Polymers}
\begin{figure}[htbp]
\begin{center}
\includegraphics[width=3in]{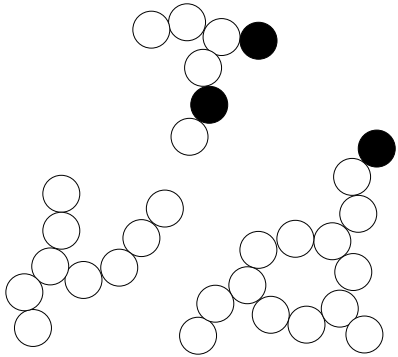}
\caption{Some of the vanishing terms in the expansion of partirion function}
\label{fig:zero-contribution}
\end{center}
\end{figure}

We begin this section by defining four new functions:
\begin{eqnarray}
P(r_{ij}^2)&=& e^{-U(r_{ij}^2)}\\
S(r_i^2)&=& e^{-V(r_i^2)}
\end{eqnarray}
Notice that if we Taylor-expand $P(R_{ij}^2)$ and $R(R_i^2)$ around $r_{ij}^2$ and $r_i^2$, because of the properties of Grassman numbers, only the first two terms survive
\begin{eqnarray}
P(R_{ij}^2)&=&P(r_{ij}^2)+\thetab_{ij}\theta{ij} Q(r_{ij}^2)\\
S(R_r^2)&=&P(r_i^2)+\thetab_i \theta_i T(r_i^2)
\label{eq:PQST}
\end{eqnarray}
In the above equations, $Q$ and $T$ are the first derivatives of the functions $P$ and $S$ respectively. Using the above equations, we can expand the grand partition function of a supersymmetric gas. We can construct a graphical expression for each term. We indicate the term $Q(r_{ij}^2)$ by a pair of connected particles (monomers) at $r_i$ ad $r_j$. The term $T(r_i^2)$ is represented by a solid black circle at point $r_i$ representing a root monomer. For each non-root monomer, represented by a hollow circle, located at $r_i$ we multiply a factor of $S(r_i^2)$. For any pair of monomers, located at $r_i$ and $r_j$, which are not connected we multiply a factor of $P(r_{ij}^2)$. For each monomer we also have a factor of $z$. The supersymmetric grand partition function is the sum of all these terms (graphs). Each term has a product of its connected graphs. It can be shown that any connected graph which has a loop or no root or even more than one root is zero (See figure \ref{fig:zero-contribution}).

Thus, only those graphs which are products of connected, loopless and single rooted graphs, ie: branched polymers, contribute. As the result of Grassman integration, for each monomer a factor of $-1/\pi$ is generated. In order to go further we need to specify the potentials $U(r_{ij}^2)$ and $V(r_i^2)$.

We choose a hardcore repulsive potential for $U$, $a$ being the dimameter of monomers. Imagine a spherical container for the particles of radius $R$. The external potential is zero inside and infinitely large outside. This results in step functions for $P$ and $S$:
$U({r}_{ij}^2) = \left\{
\begin{array}{ll}
    \infty & {r}_{ij}^2 <a^2 \\
    0 & {r}_{ij}^2 > a^2
\end{array}\right.$ $\Rightarrow$
$P({r}_{ij}^2)= \left\{
\begin{array}{ll}
    0 & {r}_{ij}^2 <a^2 \\
    1 & {r}_{ij}^2 > a^2
\end{array}\right.$

$V({r}_{i}^2) = \left\{
\begin{array}{ll}
    0 & {r}_{i}^2 <R^2 \\
    \infty & {r}_{i}^2 > R^2
\end{array}\right.$ $\Rightarrow$
$S({r}_{i}^2)= \left\{
\begin{array}{ll}
    1 & {r}_{i}^2 <R^2 \\
    \infty & {r}_{i}^2 > R^2
\end{array}\right.$
Because of this,  $Q$ and $T$ take the form of delta functions
$\begin{array}{lll}
Q({r}_{ij}^2) &=\delta({r}_{ij}^2-a^2) &=\frac{1}{2a}\delta(r_{ij}-a)\\
T({r}_{i}^2) &=\delta({r}_{i}^2-R^2) &=-\frac{1}{2R}\delta(r_{i}-R)
\end{array}$

With the above choice of functions, any pair of connected monomers have a fixed separation of $a$, the diameter of monomers. Also, any root monomer, because the delta function  is constrained to stay on the surface of the sphere. For any root monomer we have a factor of $-1/2R$ and for all other monomers a factor of $1/2a$. Recall that there is also a factor of $-z/\pi$ for any monomer. Therefore, $z/2\pi R$ appears as the fugacity of the roots and $-z/2\pi a$ as the fugacity of other monomers. Putting everything together we obtain
\begin{equation}
G_{SSG}(z)=G_{BP}(-\frac{z}{2\pi a}, \frac{z}{2\pi R})
\label{eq:ssg-bp}
\end{equation}
Comparing this with our previous equation \eqref{eq:ssg-hg} provides the main result:
\begin{equation}
G_{HG}(z)=G_{BP}(-\frac{z}{2\pi a}, \frac{z}{2\pi R})
\end{equation}

\subsection{Discussion}
 As we see the fugacities in $G_{BP}$ appear with opposite signs in the equation \eqref{eq:main-spherical}. However, the physical region of the partition function of $G_{BP}$ is where both of its fugacity arguments are positive. Consequently, one cannot explore the physical region of the branched polymer solution using this equation.

\bibliography{discrete}

\end{document}